\documentclass[aps,pra, two column,10pt,superscriptaddress,longbibliography,nofootinbib]{revtex4-2}
\usepackage{etex}
\usepackage{physics}
\usepackage{amsmath,amssymb,amsthm,amstext,mathrsfs}
\usepackage{graphicx}
\usepackage{wrapfig} 
\usepackage{tikz}
\usetikzlibrary{arrows.meta,positioning,calc,fit}
\usepackage{pgfplots}
\pgfplotsset{compat=1.18}

\usepackage[colorlinks=true,citecolor=blue,urlcolor=blue]{hyperref}
\usepackage{comment}

\usepackage{booktabs}
\usepackage{array}
\usepackage{xcolor}
\usepackage{xfrac}

\usepackage{enumerate}
\usepackage{times,txfonts}
\usepackage{multirow}
\usepackage[table]{xcolor}
\usepackage{braket}
\usepackage{svg}
\usepackage{color}
\usepackage{natbib}
\usepackage{amsmath,blkarray}
\usepackage{mathtools}
\usepackage{latexsym}
\usepackage{tabularx, booktabs}
\usepackage{graphics,epstopdf}
\usepackage{lipsum} % graphicx already loaded above
\usepackage{float}
\usepackage{amsfonts,dsfont}
\usepackage{subcaption}
\usepackage{enumerate}
\usepackage{color,soul}

\newcommand{\be}{\begin{equation}}
\newcommand{\ee}{\end{equation}}
\newcommand{\ba}{\begin{eqnarray}}
\newcommand{\ea}{\end{eqnarray}}

\newcommand{\half}{\frac{1}{2}}

\newtheorem{thm}{Theorem}
\newtheorem{Lemma}{Lemma}

\definecolor{ss}{RGB}{250,80,220}
\begin{document}

\title{Communication Supremacy of a Qubit in Holevo–Frenkel–Weiner Scenario}

\author{Partha Patra}
\email{parthapatra144@gmail.com}
\affiliation{Centre for Interdisciplinary Programs, Indian Institute of Technology Hyderabad, Telangana-502284, India}

\author{Sumit Mukherjee}
\email{mukherjeesumit93@gmail.com}
\affiliation{S. N. Bose National Center for Basic Sciences, Block JD, Sector III, Salt Lake, Kolkata 700106, India}

\author{A. K. Pan}
\email{akp@phy.iith.ac.in (Corresponding author)}
\affiliation{Department of Physics, Indian Institute of Technology Hyderabad, Telangana-502284, India}

\begin{abstract}
The prevailing wisdom asserts that in the Holevo-Frenkel-Weiner (HFW) framework of communication tasks, a qubit offers no advantage over a c-bit  unless additional quantum resources are supplied or the classical shared randomness (SR) is restricted. In contrast, we establish the supremacy of qubit communication in the HFW framework when  additional quantum resources—such as entanglement or measurement incompatibility are not injected and SR is a free resource. We consider a prepare-measure scenario where the sender, Alice, prepares six states subject to a specific cryptographic constraint, and the receiver, Bob, performs a three-outcome measurement. Such a constraint on preparation impacts the power of qubit communication quite differently than that of c-bit communication, ultimately giving rise to the quantum advantage. We introduce a witness capable of distinguishing quantum strategies that lie beyond the set of possible classical strategies and demonstrate the optimal quantum advantage. {As an application, we demonstrate the generation of certified randomness without requiring any seed randomness, since Bob has no input.}  We further analyse the robustness of the witness and show that the advantage persists in noisy situations, making it suitable for experimental realisation. Finally, we demonstrate that Bob's three-outcome qubit  positive-operator-valued-measure (POVMs) has a distinct capability of detecting coherence, which can be regarded as the non-classical feature responsible for showcasing the quantum advantage. %  the potential of this specific class of POVMs in quantum information processing.

 %We investigate the efficiency of qubit communication in constrained prepare-measure communication game within Holevo-Frenkel-Weiner framework. In the introduced communication game the senders preparations entails specific cryptographic constraint, while the receiver implements a single generalised measurement. We show that these restrictions in preparation affect the merits of c-bit and qubit communication in such a different way that the later achieves supremacy. In particular, we prove that qubit communication can outperform any classical bit strategy even when unlimited shared randomness is available. Moreover, we construct an explicit example of quantum encoding decoding scheme that achieves maximum advantage compared to the c-bit communication without requiring additional quantum resources such as incompatibility or entanglement. We further analyse the robustness of protocol against white noise and show that the advantage persists considerably under noisy conditions making it suitable for experimental realisation. Finally, we identify POVM-based coherence as essential nonclassical resource responsible for the observed quantum advantage, highlighting the potential of this particular class of POVM in shaping the future of quantum information processing.
\end{abstract}
 
\maketitle

\section{Introduction}
The advancement of global communication has traditionally relied on the optimisation of classical  channels, wherein information is encoded into discrete binary digits (c-bits) and transmitted through distinguishable physical states. However, the Shannon–Hartley theorem \cite{Shannon1948,Shannon1949,Hartley1928Transmission} imposes a fundamental upper bound on the information-carrying capacity of such channels and eventually inspires the exploration of quantum information theory \cite{Wilde2017,NielsenChuang2010} as an alternative  communication paradigm. In this framework, the qubit replaces the c-bit, extending the discrete state space $\{0,1\}$ to a continuous manifold of superposed states. Although this seemingly allows for infinitely many encoding possibilities, the amount of accessible information is fundamentally constrained by the celebrated Holevo theorem \cite{Holevo1973}.

In a general HFW communication scenario, Alice and Bob are two parties where, for each input $x\in \mathcal{X}$, the sender Alice, following a pre-decided strategy, prepares a state $\mathbb{P}_x$  according to the probability distribution $p(x)$ and communicates it to the receiver Bob. Bob has no choice of measurement, \emph{i.e.}, he performs a fixed measurement and outputs  $b\in \mathcal{B}$ such that the prediction probability of a given functional $f(x,b)$ is maximized. Now, Alice may choose a quantum channel over a classical channel to send information. The seminal result by Holevo \cite{Holevo1973} asserts that allowing Alice to use a $d$-level quantum system (qudit) for state preparation does not offer any advantage over a $d$-level classical system (c-dit). More precisely, the channel capacity quantified by the maximum mutual information $\max_{\mathbb{P}_x}\mathcal{I}(x:b)$ is the same for quantum and classical channels. Therefore, the extractable information is same in both cases. Subsequently, Frenkel and Weiner \cite{FW_2015} strengthened this result by demonstrating that, in the same scenario, any arbitrary $f(x,b)$ constructed by the operational statistics of any qudit can be reproduced by c-dit, provided that the SR between the parties is treated as a free resource.

However, such an equivalence in the HFW scenario can be broken either by allowing additional quantum resources, such as entanglement and measurement incompatibility, or by limiting the SR. For instance,  preshared entanglement between Alice and Bob offer a quantum advantage, as demonstrated in the dense coding protocol \cite{densed_coding_92} and can also enhance capacity of classical channel \cite{Frenkel2022entanglement} as well. Likewise,  quantum supremacy can be revealed by introducing seed randomness into Bob’s input in communication games\cite{Wiesner1983RAC,Ambainis1999RAC,Ambainis2002RAC,abhyoudai23,Pan2019} where measurement incompatibility of Bob’s decoding measurements serves as a necessary resource for extracting the quantum advantage \cite{Carmeli_2020, saha2023pra}. On the other hand, instead of allowing additional quantum resources, restricting the SR can also reveal quantum supremacy \cite{Patra2024,chen_banik2024}. However the above scenarios go beyond the strict HFW framework. Therefore, the common consensus is that the quantum advantage must arise either from additional quantum resources or from limiting classical correlations (SR) because compatible measurements reduce Bob’s measurement to a single grand POVM \cite{busch1996quantum}, and unbounded SR combined with c-dit communication can reproduce any q-dit statistics\cite{FW_2015}.  

%Against this backdrop, we raise an important question as follows. If one forbids measurement incompatibility and entanglement while allowing unbounded SR, can a qubit channel demonstrate a communication advantage over a c-bit channel? At first glance it seems to be impossible because compatible measurements reduce Bob’s device to a single grand POVM \cite{busch1996quantum}, and unbounded SR combined with c-dit communication can reproduce any q-dit statistics\cite{FW_2015}. Therefore, the common consensus is that the quantum advantage must arise either from additional quantum resources or from limiting classical correlations (SR). The central result of this work  shows that this intuition is incomplete. 

In contrast to the common consensus, we demonstrate that qubit communication outperforms any c-bit communication strategy in the HFW scenario without invoking any additional quantum resources and allowing  unbounded SR between the parties. We introduce a prepare-measure communication game in which the sender, Alice, prepares six states, and the  receiver, Bob, performs a single three-outcome measurement.  Alice's preparation is subjected to an operational restriction, namely, the parity concealment constraint. We define a suitable witness—the success probability of the game, and demonstrate the communication supremacy of a qubit over a c-bit by fully characterizing the classical correlation polytope under this constraint. Further, we show that our communication game can be utilised to certify intrinsic quantum randomness in the prepare-measure scenario without requiring seed randomness.

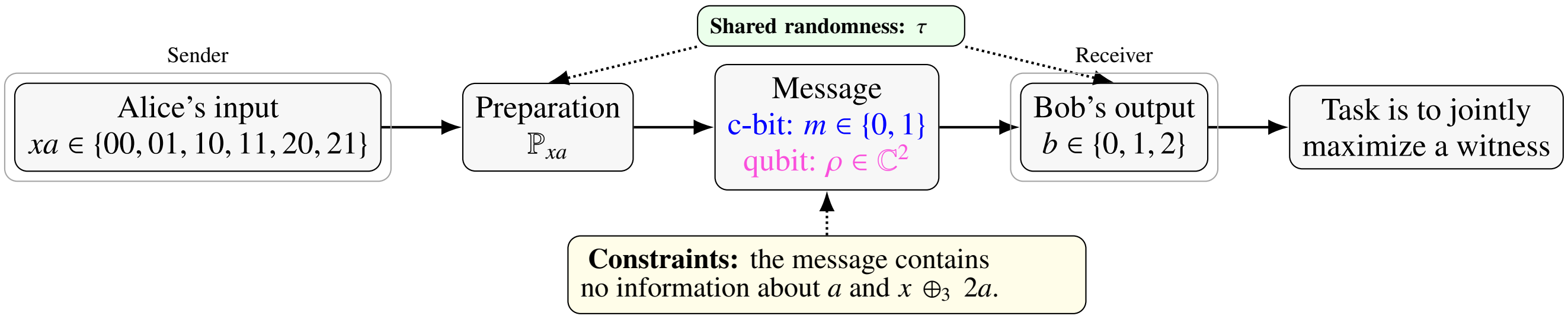
\begin{figure*}[t]
\centering
\begin{tikzpicture}[
  font=\normalsize,
  node distance=6mm and 9mm,
  box/.style={draw,rounded corners,align=center,inner sep=4pt,minimum height=8mm},
  stage/.style={box,fill=black!3},
  arr/.style={-Latex,thick},
  note/.style={draw,rounded corners,align=left,inner sep=4pt,font=\scriptsize,fill=yellow!10}
]

\node[stage] (alice) {Alice's input\\  {${xa}\in \{00,01,10,11,20,21\}$}};
\node[stage, right=of alice] (prep) {Preparation\\ $\mathbb{P}_{xa}$};
\node[stage, right=of prep] (chan) {Message\\\textcolor{blue}{c-bit: $m\in\{0,1\}$}\\\textcolor{ss}{qubit: $\rho\in\mathbb{C}^2$}};
\node[stage, right=of chan] (bob) {Bob's output\\ $b\in\{0,1,2\}$};
\node[stage, right=of bob] (win) {Task is to jointly\\ maximize a witness};

\draw[arr] (alice) -- (prep);
\draw[arr] (prep) --  (chan);
\draw[arr] (chan) -- (bob);
\draw[arr] (bob) -- (win);

\node[note, above=9mm of $(prep)!0.5!(bob)$, text width=27mm, fill=green!8] (sr)
{\textbf{Shared randomness:} $\tau$};
\draw[-Latex,thick,densely dotted] (sr.south west) -- (prep.north);
\draw[-Latex,thick,densely dotted] (sr.south east) -- (bob.north);

\node[note, below=5mm of chan, text width=55mm] (constraint)
{ \small{ \textbf{Constraints:} the message contains\\ no information about $a$ and $x\oplus_3 2a$.}};
\draw[-Latex,thick,densely dotted] (constraint.north) -- (chan.south);

\node[fit=(alice),draw=black!35,rounded corners,inner sep=3pt,label={[font=\scriptsize]above:Sender}] {};
\node[fit=(bob),draw=black!35,rounded corners,inner sep=3pt,label={[font=\scriptsize]above:Receiver}] {};

\end{tikzpicture}
\caption{A schematic diagram for prepare measure communication game in HFW scenario subjected to parity-concealment constraints.}
\label{fig:scenario_schematic}
\end{figure*}

 We also explore the experimental realisability of our witness by analyzing its robustness under white noise. We show that the obtained quantum advantage remains robust even in the presence of a substantial amount of noise, as the gap between the classical and quantum values of the witness is sufficiently large.  We note that quantum supremacy is revealed in the absence of entanglement and measurement incompatibility. It is then curious to probe what could be the non-classical feature responsible for enhancing such information processing capability in the HFW scenario. We identify that Bob's three-outcome qubit POVM has a special power to detect coherence, leading to the quantum advantage. Specifically, we demonstrate that whenever the qubit strategy gives an advantage,  the POVM-based coherence measure is nonzero. % it is necessary for such enhancement. The combined effect of POVM-based coherence and the parity concealment constraint enables qubit advantage that surpasses its classical counterpart.

\section{{Results}} 

\emph{Scenario —} Consider a prepare-measure communication game between Alice and Bob where Alice receives one of the six uniformly random input messages $\{xa\}\in \{00,01,10,11,20,21\}$, with $x\in\{0,1,2\}$ and $a\in \{0,1\}$, encodes the message by preparing either a classical or a quantum system by corresponding preparation procedure $\mathbb{P}_{xa}$, and sends to Bob. Subsequently, Bob performs a single fixed measurement $\mathbb{M}$ that produces three outputs $b\in \{0,1,2\}$. The task of the parties is to jointly evaluate an arbitrary function $\mathcal{F}(b,a,x)$.

Alice can communicate either a quantum or a classical message ($m$) of the same dimension to Bob. In this work, we focus only on the two dimensional system. Furthermore, Alice's communication abides by specific cryptographic restrictions inspired by \cite{Spekkens2009,ghorai2018}, so that the message contains no information about $a$ and $x\oplus_3 2a$ ($\oplus_3$ denotes addition modulo $3$). We refer to these restrictions as \textit{parity concealment constraints}, which can be interpreted as operational privacy requirements on the communication. %In other words, no information about $a$ and no information about $x\oplus_3 2a$ can be carried by the communicated system. 
Mathematically, the constraints can be expressed as follows.

{\small
\begin{subequations}
\begin{align}
  \forall m\quad &\sum_{x} p\bigl(m\mid\mathbb{P}_{x0}\bigr)
  = \sum_{x} p\bigl(m\mid\mathbb{P}_{x1}\bigr),
  \label{Parity_Constraint}\\
  \forall m\quad &\sum_{\substack{xa: x\oplus_3 2a=0}} p\bigl(m\mid\mathbb{P}_{xa}\bigr)
  = \sum_{\substack{xa: x\oplus_3 2a=1}} p\bigl(m\mid\mathbb{P}_{xa}\bigr)
  = \sum_{\substack{xa:x\oplus_3 2a=2}} p\bigl(m\mid\mathbb{P}_{xa}\bigr)
  \label{Parity_Constraint2}
\end{align}
\end{subequations}}
where, $x\in\{0,1,2\}$ and $a\in \{0,1\}$. They may utilize unlimited shared randomness ($\tau$) to design strategies that optimize performance for a specified task. Note, however, that $\tau$ is directly accessible to both parties. Consequently, Alice must encode the information in a conveying message $m$ to Bob such that he, even with access to the shared randomness $\tau$ and all available decoding resources, cannot extract any parity information from the message. This requires that the parity-concealment constraints in Eqs.~\eqref{Parity_Constraint} and \eqref{Parity_Constraint2} hold for every fixed value of \(\tau\), and not merely for the probability distribution averaged over $\tau$.

 Now, every strategy that can be implemented by the parties to evaluate the function can be represented by a \textit{channel matrix} $P \in \mathscr{C} \subset \mathbb{R}^{3\times6}$ consists of a collection of conditional probabilities $\{p(b|x,a)\}_{b,xa}$ and is given by \cite{FW_2015},
\begin{equation}
  P:=  \big\{p(b|x,a)\big\}_{b,xa},  \ b\in \{0,1,2\}, \ xa\in \{00,01,10,11,20,21\}
\end{equation}
where $\sum_{b}p(b|x,a)=1$ and therefore the channel matrix is column stochastic.

%\textcolor{red}{Obviously, there exists a strategy for evaluating an arbitrary function $\mathcal{F}(b,a,x)$ using either a three-level classical system or a qutrit that realises the ideal perfect channel $P_{perfect}$; however, such a strategy could violate the parity concealment restrictions given by Eqs. \eqref{Parity_Constraint} and \eqref{Parity_Constraint2}. This introduces a nontrivial feature to the communication game by compelling us to restrict the scenario to one where Alice can only send a two-level system. However, using a two-level system, whether classical or quantum a perfect channel cannot be implemented. This naturally raises the following question: Does there exist a strategy, satisfying the parity concealment constraints \eqref{Parity_Constraint} and \eqref{Parity_Constraint2}, that can be implemented with a qubit but whose corresponding channel matrix cannot be reproduced by any strategy employing a classical two-level system? In the following we will answer this question affirmatively.}

Our study focuses on such communication scenarios subjected to such cryptographic constraints as Eqs. \eqref{Parity_Constraint} and \eqref{Parity_Constraint2} and we aim to investigate whether the equivalence between classical and quantum channel matrices appears in HFW framework remains intact.

We first consider the classical scenario of one c-bit communication. For the classical channel $P_C:=\{p(b|x,a)\}_{b,xa} \in \mathcal{C}$ with a c-bit communication, we have 

\begin{equation}\label{eq:cl_channel}
    p(b|x,a)= \sum_{m\in\{0,1\}} p(m|x,a) p(b|m),
\end{equation}
where $p(m|x,a)$ and $p(b|m)$ depends on the encoding and decoding strategy respectively with $m$ being the one c-bit message. In the most general scenario, the parties have a pre-decided encoding scheme ($\mathds{E}_1$) and decoding scheme ($\mathds{D}_1$) as specified below.\\

\textbf{Encoding strategy ($\mathds{E}_1$)}: For the input $xa$, Alice prepares
 `$0$' with probability $p_{xa}$ and `$1$' with probability $1-p_{xa}$, where $p_{xa}=p\left(m=0|x,a\right)$.\\

 \textbf{Decoding strategy ($\mathds{D}_1$)}: If Bob receives `$0$' he outputs $0$ with probability $s_0$, $1$ with probability $s_1$ and $2$ with probability $s_2$. Similarly, if he receives `$1$', he outputs $b$ with probability $r_b, \ \forall b\in \{0,1,2\}$ with $\sum_b s_b=\sum_b r_b=1$. Here, $s_b=p(b|0)$ and $r_b=p(b|1)$. An explicit description of the strategy is given by, 
\begin{center}
\resizebox{\columnwidth}{!}{
\begin{tabular}{||c| c c c c c c| c ||} 
 \hline
 Message & 00 & 01 & 10 & 11 & 20 & 21 & Decoding \\ [0.5ex] 
 \hline\hline
 0 & $p_{00}$ & $p_{01}$ & $p_{10}$ &$p_{11}$ & $p_{20}$ & $p_{21}$ & $s_0\rightarrow0$;\ $s_1 \rightarrow 1$;\ $s_2\rightarrow 2$ \\ 
 \hline
 1 & 1-$p_{00}$ & 1-$p_{01}$ & 1-$p_{10}$ & 1-$p_{11}$ & 1-$p_{20}$ & 1-$p_{21}$ & $r_0\rightarrow0$;\ $r_1 \rightarrow 1$;\ $r_2\rightarrow 2$ \\[1ex] 
 \hline 
\end{tabular}}
\end{center}

In order to satisfy the restriction \eqref{Parity_Constraint} and \eqref{Parity_Constraint2} of the game, Alice's preparations adhere,

\begin{subequations}
 \begin{align}\label{PO_c1}
&\frac{1}{3}\left(p_{00}+p_{10}+p_{20}\right)= \frac{1}{3}\left(p_{01}+p_{11}+p_{21}\right)=\kappa_1\\
\label{PO_c2}
&\frac{1}{2}\left(p_{00}+p_{11}\right)=\frac{1}{2}\left(p_{01}+p_{20}\right)=\frac{1}{2}\left(p_{10}+p_{21}\right)=\kappa_2
\end{align}
\end{subequations}
where $\kappa_1, \kappa_2\in [0,1]$.
According to the aforementioned most general encoding and decoding strategy, classical channel matrix can be written as,

\begin{widetext}
\begin{equation}
\resizebox{\columnwidth}{!}{$
P_{C}=
\begin{pmatrix}
 p_{00}s_0+ (1-p_{00})r_0& p_{01}s_0+ (1-p_{01})r_0 &p_{10}s_0+ (1-p_{10})r_0 & p_{11}s_0+ (1-p_{11})r_0 &p_{20}s_0+ (1-p_{20})r_0 &p_{21}s_0+ (1-p_{21})r_0 \\
 p_{00}s_1+ (1-p_{00})r_1& p_{01}s_1+ (1-p_{01})r_1 &p_{10}s_1+ (1-p_{10})r_1 & p_{11}s_1+ (1-p_{11})r_1 &p_{20}s_1+ (1-p_{20})r_1 &p_{21}s_1+ (1-p_{21})r_1  \\
 p_{00}s_2+ (1-p_{00})r_2& p_{01}s_2+ (1-p_{01})r_2 &p_{10}s_2+ (1-p_{10})r_2 & p_{11}s_2+ (1-p_{11})r_2 &p_{20}s_2+ (1-p_{20})r_2 &p_{21}s_2+ (1-p_{21})r_2
\end{pmatrix}
$}\label{Classical_channel_Mat}
\end{equation}
\end{widetext}

On the other hand, while using a quantum strategy Alice prepares six qubit states $\{\rho_{xa}\}$ corresponding to the preparation procedures $\{\mathbb{P}_{xa}\}$. Upon receiving the qubit from Alice, Bob implements the three-outcome qubit measurement $\mathbb{M}\equiv\{E_b | \ E_b=\alpha_{b} \pi_{b}, \ \sum_{b=0}^{2}E_{b}= \mathds{I} \}$.
In this case, the elements of the channel matrix $P_{Q} \in \mathcal{Q}$ corresponding to a particular quantum strategy is determined by probabilities derived from Born’s rule.  Such a matrix can be written as follows, 
\begin{equation}
\resizebox{0.95\columnwidth}{!}{$
P_{Q} =
\begin{pmatrix}
 \Tr[\rho_{00} E_0] & \Tr[\rho_{01} E_0] & \Tr[\rho_{10} E_0] & \Tr[\rho_{11} E_0] & \Tr[\rho_{20} E_0] & \Tr[\rho_{21} E_0] \\
 \Tr[\rho_{00} E_1] & \Tr[\rho_{01} E_1] & \Tr[\rho_{10} E_1] & \Tr[\rho_{11} E_1] & \Tr[\rho_{20} E_1] & \Tr[\rho_{21} E_1] \\
 \Tr[\rho_{00} E_2] & \Tr[\rho_{01} E_2] & \Tr[\rho_{10} E_2] & \Tr[\rho_{11} E_2] & \Tr[\rho_{20} E_2] & \Tr[\rho_{21} E_2]
\end{pmatrix}
$}\label{quantum_channel_Mat}
\end{equation}
where $\mathcal{Q}$ is the set of all channel matrices that can be constructed by all possible quantum strategies.
Note that, to satisfy Eqs. (\ref{Parity_Constraint}) and (\ref{Parity_Constraint2}) the communicated qubit states must satisfy,
\begin{subequations}
    \begin{eqnarray}\label{Q_PO_Condition1}
   &\sum\limits_{x\in\{0,1,2\}}\rho_{x0} =\sum\limits_{x\in\{0,1,2\}}\rho_{x1}& \\
  & \label{Q_PO_Condition2} \sum\limits_{xa:x\oplus_32a=0}\rho_{xa} =\sum\limits_{xa:x\oplus_3 2a=1}\rho_{xa}=\sum\limits_{xa:x\oplus_3 2a=2}\rho_{xa}&
\end{eqnarray}
\end{subequations}

We are in a position to prove our main result which is captured by the following theorem.

\begin{thm}\label{ThmC<Q}
    For the communication task in the HFW scenario augmented by a parity concealment constraint, the set of classically realisable channels constitutes a strict subset of those attainable with quantum strategies, i.e., $\mathcal{C} \subsetneq \mathcal{Q}$, where $\mathcal{C}$ and $\mathcal{Q}$ respectively denote the sets of all classical and quantum channel matrices.
\end{thm}

\begin{proof} This proof consists of two parts. First, we show that any channel matrix arising from a c-bit communication can also be realised using qubit communication, i.e., $\mathcal{C} \subseteq \mathcal{Q}$. Second, we show that not every qubit channel matrix can be simulated by classical c-bit communication so that $\mathcal{C} \neq \mathcal{Q}$. Consequently, demonstrating the existence of a single such quantum channel is sufficient to complete the proof.

To prove the first part, we note that given any $P_C$ of the form Eq.~\eqref{eq:cl_channel} there always exists a qubit channel $P_{Q}$ that simulates $P_{C}$ and vice-versa \cite{FW_2015}. To see this let us consider the following qubit states and measurements as,
\begin{eqnarray}\label{eq:quRealisation}
  &&\rho_{xa}=p(0|x,a)\ketbra{0}{0}+ p(1|x,a)\ketbra{1}{1} \nonumber \\
 && E_{b}=p(b|0)\ketbra{0}{0}+p(b|1)\ketbra{1}{1},
\end{eqnarray}
where $\{E_{b}\}$ is a  POVM with $\sum_{b}E_{b}=\mathbb{I}$ since $\sum_{b}p(b|m)=1 \ \forall m \in \{0,1\}$. This choice exactly reproduce Eq.~\eqref{eq:cl_channel}.

 However, the crucial point is to determine whether, upon imposing the {parity concealment constraints} in Eqs. \eqref{PO_c1} and \eqref{PO_c2}, all c-bit channels can be simulated by qubit channels. 
 
 Since, according to Eq.~\eqref{PO_c1} the channel $P_C$ in Eq.~\eqref{eq:cl_channel} must satisfy
\begin{equation}
   \sum_x p(b|x,0) = \sum_x p(b|x,1) \quad \forall b \in \{0,1,2\},
\end{equation}

then, $\rho_{xa}$ in Eq.~\eqref{eq:quRealisation} also obey the Eq.~\eqref{Q_PO_Condition1}, as $\sum_x \rho_{x0}=\sum_x \rho_{x1}$. The same argument applies to constraint Eq.~\eqref{Parity_Constraint2} as well.

The set of classical channel matrices $\mathcal{C}$  constitute a convex set. Consequently,  any classical strategy assisted by SR is another classical strategy and is given by,

\begin{equation}
    p(b|x,a)=\sum_{\tau}p(\tau)p_{\tau}(b|x,a),
\end{equation}
where, each $p_{\tau}(b|x,a)$ takes the form of Eq.~\eqref{eq:cl_channel}, with $\tau$ being the shared random variable. 
Since each $p_{\tau}(b|x,a)$  can be simulated using qubits, the overall distribution $p(b|x,a)$ is likewise simulable when the qubit strategy that involves SR.
%Therefore, it suffices to prove that any classical channel corresponding to deterministic strategies (with fixed $\tau$) can be simulated by a quantum channel. 

Then for any $P_C\in \mathcal{C}$  have a quantum realisation of Eq.~\eqref{eq:quRealisation} which automatically satisfies constraints in Eqs.~\eqref{Parity_Constraint} and \eqref{Parity_Constraint2} if its classical counterpart do so and therefore, $\mathcal{C} \subseteq \mathcal{Q}$.
%The first part is straightforward: for the classical bits $\{0,1\}$ we consider the corresponding quantum states $\{\ket{0},\ket{1}\}$. Bob performs measurements in the $\sigma_z$ basis. This setup reproduces exactly the same channel matrix $\mathcal{C}$ as in Eq.~\eqref{Classical_channel_Mat}, $i.e.$, $\mathcal{C}\subseteq\mathcal{Q}$. 

%Alternatively, one may argue that the parties can use pre-shared classical correlations $p(\tau)$ via a shared random variable $\tau$. They can then adopt various strategies and, conditioned on the value of $\tau$, implement the strategy $\mathcal{C}(\tau)$. Consequently, the most general classical channel can be written as the matrix
%$\mathcal{C}_{SR} = \sum_{\tau} p(\tau)\,\mathcal{C}(\tau)$. The question is whether every such classically generated strategy lies within the quantum set. The answer is affirmative because in quantum case also they would the same shared randomness and obtain the same channel matrix as $\mathcal{Q}_{SR}=\sum_{\tau} p(\tau)\,\mathcal{Q}(\tau)$, therefore all possible convex hull of classical strategy can be simulated by quantum strategy, $i.e.$, $\mathcal{C}_{SR}\subseteq\mathcal{Q}_{SR}$. 

Now, to prove the second part, we construct a witness that can distinguish classical from quantum strategies. For this, we introduce a communication game in the aforementioned scenario, in which the parties win in a particular run if Bob outputs $b = x \oplus_3 a$ for Alice's input $xa$. Technically, the function that the parties jointly evaluate is
\begin{align}\label{eq:function}
\mathcal{F}(b,a,x) = \begin{cases}
1 & \text{for } b = x \oplus_3 a\\[2pt]
0 &  \text{otherwise}\\[2pt]
\end{cases}
\end{align}

We define the witness as the average success probability that evaluates the function in Eq.~\eqref{eq:function} as follows.
\begin{eqnarray}\label{Avg_Success}
    \mathcal{P}= \frac{1}{6} \sum_{{x\in \{0,1,2\}}\atop a\in\{0,1\}} \sum_{b\in \{0,1,2\}}\mathcal{F}(b,a,x)p(b|\mathbb{P}_{xa}).
\end{eqnarray}

To demonstrate the existence of qubit channels that cannot be simulated by any c-bit channel, we present the following Lemma.
\begin{Lemma}
    \label{qubit>cbit}
 (i) The maximum success probability achievable by a c-bit communication in the presence of unbounded SR, is $\mathcal{P}_{C}^{max}=7/12$. (ii) There exists a qubit channel matrix $P_Q$ that gives the optimal quantum success probability $\mathcal{P}_{Q}^{max}=\frac{1}{3}\left(1 + \frac{\sqrt{3}}{2}\right)$.
\end{Lemma}
\begin{proof}
The proof of (i) is quite lengthy and hence deferred to the Appendix-\ref{optimisation of bit comm}. We provide the sketch of the proof here. We start from the general channel matrix $P_C$ in Eq.~\eqref{Classical_channel_Mat} and, by considering unbounded SR among admissible strategies, compute the average success probability using Eq.~\eqref{Avg_Success}. Imposing the constraints Eqs.~(\ref{PO_c1}) and \eqref{PO_c2}, we then maximize the average success probability and achieve the maximum value of $\mathcal{P}_C^{max}=7/12$.

{To provide an alternative and elegant proof of (i), we provide a complete characterization of the set of classical correlations in Appendix-\ref{Ap:Polytope}. The set of all possible classical strategies that abide by the parity concealment constraints in Eq.~\eqref{Parity_Constraint} and  \eqref{Parity_Constraint2},  constitutes a six-dimensional polytope with $21$ vertices described in Table~\ref{Tab:21_vertices} . When shared randomness $\tau$ is allowed, any point inside this polytope can be obtained as a convex combination of these vertices. We further establish that the maximum value achievable by our witness Eq.~\eqref{Avg_Success} within this polytope, is $\mathcal{P}_C^{\max} = \frac{7}{12}$.}

To prove (ii), we use specific quantum encoding–decoding scheme to analyse the performance of qubit communication in terms of the average success probability in Eq.~\eqref{Avg_Success} and prove that no classical strategy can achieve the corresponding value of $\mathcal{P}_Q$. For this, consider the  following quantum channel matrix, 
\begin{equation}
P_\mathcal{Q}=
\begin{pmatrix}\label{Q_Max_Channelmatrix}
\frac{2+\sqrt3}{6} & \frac{1}{3}         & \frac{2-\sqrt3}{6} & \frac{2-\sqrt3}{6} & \frac{1}{3}         & \frac{2+\sqrt3}{6}\\
\frac{1}{3}         & \frac{2+\sqrt3}{6} & \frac{2+\sqrt3}{6} & \frac{1}{3}         & \frac{2-\sqrt3}{6} & \frac{2-\sqrt3}{6}\\
\frac{2-\sqrt3}{6} & \frac{2-\sqrt3}{6} & \frac{1}{3}         & \frac{2+\sqrt3}{6} & \frac{2+\sqrt3}{6} & \frac{1}{3}
\end{pmatrix}
\end{equation} 

which can be realised by the following strategy. For $a=0$, Alice prepares the qubit trine-states
\begin{equation}
\rho_{x0}=\ketbra{\phi_x}{\phi_x}=\frac{1}{2}\left(\mathds{I}+\cos\theta_x\,\sigma_Z+\sin\theta_x\,\sigma_X\right),\ \forall x\in\{0,1,2\},
\end{equation}

where, $\theta_x=\frac{2\pi x}{3}$. 
For $a=1$, the preparations $\rho_{x1}$ are orthogonal to $\rho_{x0}$, so that,
\begin{equation}
\rho_{x1}=\ketbra{\bar{\phi}_{x\oplus_3 2}}{\bar{\phi}_{x\oplus_3 2}} \ \forall x\in\{0,1,2\}.
\end{equation}

It is straightforward to check  that the constraints on preparations in Eqs.~(\ref{Q_PO_Condition1}) and (\ref{Q_PO_Condition2}) are  satisfied for this set of preparations. 

Bob measures the trine-spin POVM, $E_b=\frac{2}{3}\ketbra{\psi_b}{\psi_b}$, with
\begin{equation}
\ketbra{\psi_b}{\psi_b}=\frac{1}{2}\left(\mathds{I}+\cos\theta_b\,\sigma_Z-\sin\theta_b\,\sigma_X\right) ,\ \forall b\in\{0,1,2\},
\end{equation}
where, $\theta_b=\frac{2\pi}{3}\left(\frac{1}{4}-b\right)$.

\begin{figure*}[t]
\centering
\begin{subfigure}[t]{0.48\textwidth}
  \centering
  \includegraphics[height=50mm, width=80mm,scale=1.5]{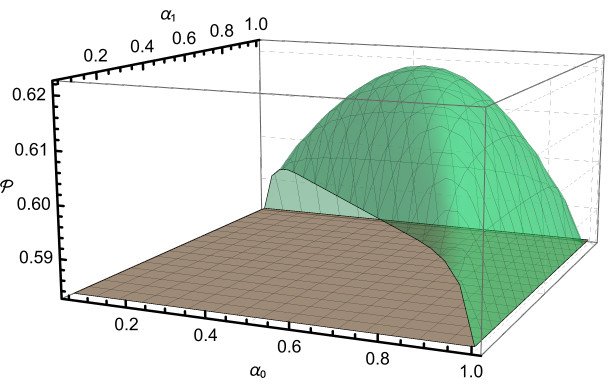}
  \caption{}
  \label{Q-C}
\end{subfigure}\hfill
\begin{subfigure}[t]{0.48\textwidth}
  \centering
  \includegraphics[height=50mm, width=85mm,scale=1.5]{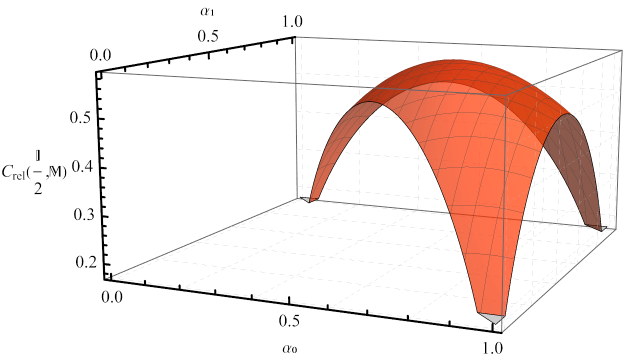}
  \caption{}
  \label{Coherence}
\end{subfigure}
\caption{ (a) Left: The success probabilities $\mathcal{P}_Q$ for qubit(green) and $\mathcal{P}_{C}$ for c-bit (gray) communication strategy are plotted against the two free parameters $\alpha_0$ and $\alpha_1$. (b) Right: POVM-based coherence $C_{rel}\left(\frac{\mathbb{I}}{2},\mathbb{M}\right)$ is plotted against $\alpha_0$ and $\alpha_1$. }
\end{figure*}

For the channel $P_\mathcal{Q}$ in Eq.~\eqref{Q_Max_Channelmatrix}, we compute the average quantum success probability in Eq.~\eqref{Avg_Success} as 
\begin{eqnarray}\label{Avg_Succ_Quantum}
    \mathcal{P}_Q\equiv\frac{1}{6}\sum_{x\in\{0,1,2\}\atop a\in\{0,1\}}\Tr[\rho_{xa} E_{x\oplus_3 a}]=\frac{1}{3}\left(1 + \frac{\sqrt{3}}{2}\right)\approx0.62
\end{eqnarray}
which is greater than $\mathcal{P}_C^{max}=\frac{7}{12}\approx 0.58$. In fact, the qubit strategy described above yields the maximal quantum advantage for  the communication game, as we prove in detail in Appendix~\ref{Appx:qo}. We also have further shown in Figure~\ref{Q-C} that for a qubit strategy that uses POVM $\{E_b=\alpha_{b}\ketbra{\psi_b}{\psi_b}\}$ with $\alpha_{b} \in [0,1]$ the success probability surpasses $\mathcal{P}_C^{max}=7/12$ for non-zero value of $\alpha_b$. This proves Lemma \ref{qubit>cbit}.

\end{proof} 

Naturally, qubit communication strategies corresponding to any channel matrix $P_\mathcal{Q}$ including one in Eq.~\eqref{Q_Max_Channelmatrix} cannot be an element of the set $\mathcal{C}$ of classical channel matrices. This implies $\mathcal{C} \subsetneq \mathcal{Q}$ in the HFW scenario. This concludes the proof of Theorem \ref{ThmC<Q}.
\end{proof}

{We provide an explicit example corresponds to one of the vertices $w_3=(1,0,\half,\half,0,1)$ from Table~\ref{Tab:21_vertices}  showing an encoding-decoding scheme that saturates the derived bound for c-bit strategy in Table \ref{tab:game}, Appendix-\ref{ced}. The self-explanatory table presents Alice’s classical message $m$ (one $c$-bit) sent to Bob, along with the corresponding output $b$ that Bob produces for a given value of the $c$-bit, corresponding to each specific input message $xa$. It can be seen from Table \ref{tab:game} that, $7$ out of the $12$ possible instances, their mutual strategy succeeds, yielding an average success probability of $\frac{7}{12}$. Note that, our witness in Eq.~\eqref{Avg_Success} is linear, and therefore it attains its maximum at one of the vertices. Nonetheless, one may propose nonlinear witnesses whose maximum may be achieved at a point in the interior of the polytope. To obtain such a point, one has to choose appropriate convex combination of the vertices using the shared random variable $\tau$.}   %Note that this strategy is not unique; multiple strategies can achieve the same success probability, reproducing the analytically derived maximum value.

%{\color{red}\emph{Remark-1}:- An immediate question could be the following. Can the quantum advantage be demonstrated in the entanglement-assisted $c$-bit communication scenario, instead of qubit communication? The answer to this question is affirmative and is quite straightforward to prove.  Alice performs three projective measurements on her shared qubit and, depending on the outcome, sends one $c$-bit so that the steered states to Bob follow the {parity concealment constraints}. Using this $c$-bit, Bob chooses to flip or not to flip his shared qubit according to their predetermined strategy and then performs the three-outcome qubit measurement. In this way, they can obtain the same success probability as $\mathcal{P}_Q^{max}$.}

\section{Robustness of quantum advantage against Noise} To analyse the robustness of the quantum advantage in the presence of noise we consider quantum states mixed with white noise \cite{vidal99}. This corresponds to writing the effective state $\rho^{noisy}$ by a convex combination of the ideal state and the maximally mixed state, so that $\rho^{noisy}_{xa}=v\rho_{xa}+(1-v)\frac{\textbf{I}}{2}$ with $v\in [0,1]$ being the visibility parameter. Operationally, this leads to a noisy success probability of the form
\begin{equation}
\label{eq:vis}
\mathcal{P}_Q(v) = v \mathcal{P}_Q^{\max} + (1-v)\frac{1}{3},
\end{equation}
where $1/3$ is the success probability obtained from completely random outcomes. Now, substituting the optimal quantum value $\mathcal{P}_Q^{\max}$ and $\mathcal{P}_C^{\max}=\mathcal{P}_Q(v) = \frac{7}{12}$ in Eq.~\eqref{eq:vis} we obtain that the quantum advantage persists with the threshold visibility $v_{th}=\frac{\sqrt{3}}{2} \approx 0.866$.

However, instead of solving explicitly for the threshold visibility, a substitute and widely used way to quantify robustness is via the \emph{white noise robustness} measure defined as the minimum amount of completely random probability has to be mixed such that the noisy quantum strategy can no longer outperform the optimal classical strategy. This is conveniently expressed by a minimisation problem as follows \cite{Baek_2025},

\begin{equation}\label{eq:wn}
R_{\mathrm{WN}} =   \min\Big\{r\geq 0 \ \Big\rvert \ \frac{\mathcal{P}_Q^{\max}+r \mathcal{P}_{N}}{1+r} \leq \mathcal{P}_C^{\max} \Big\},
\end{equation}
where $\mathcal{P}_{N}$ is the success probability obtained from completely mixed state. In the present scenario, we have $\mathcal{P}_{N}=1/3$, for Eq.~\eqref{eq:wn} yields $R_{\mathrm{WN}} \approx 0.156$ and the noise fraction $\frac{r}{1+r}=0.135$. 
Therefore, mixing $13.5\% $ of white noise into the quantum state destroys the quantum advantage, and hence the witness is robust against significant amounts of noise.

{
\section{Randomness Certification}

We demonstrate that the optimal quantum advantage in the HFW scenario can be exploited to generate certified randomness. Note that certified randomness generation in the standard prepare-measure scenario \cite{Bohr_2022,RochiCarceller_2026} involves multiple inputs and therefore requires seed randomness. However, in the HFW scenario, the input is fixed, so the resulting randomness is generated without any seed randomness. A recent work shows that seedless randomness certification is possible in a network scenario, but that  relies on entanglement \cite{Sarkar2024networkquantum}.

%The witness, defined in terms of the success probability of the communication game introduced in this work, separates quantum strategies from all possible classical strategies assisted by shared randomness. This separation naturally raises the question of whether the observed quantum advantage can be exploited to certify the intrinsic randomness of the resulting quantum statistics, in the same spirit that randomness is certified from Bell inequality violations in spatially separated experiments \cite{acin_Randomness16,YiLi_2025_RandomnessIncompatibility}, prepare-and-measure scenarios \cite{Bohr_2022,RochiCarceller_2026}, and the quantum statistics of temporally separated measurements \cite{subhayan_LGI_randomness25}. In particular, since the communication game exhibits a quantum advantage using only a single measurement, it opens the possibility of certifying randomness without requiring seed randomness, similar to what has recently been demonstrated in network scenarios \cite{Sarkar2024networkquantum}.

To quantify the amount of certifiable randomness from the quantum statistics obtained in our communication game, we consider an adversary Eve whose objective is to correctly predict Bob's measurement outcome. This randomness is characterized by Eve's optimal guessing probability, defined as the maximum probability with which she can correctly guess Bob's output $b$ over all possible quantum strategies consistent with the observed communication statistics. In our scenario, this guessing probability is given by,
\begin{equation}
    P_{guess}=\max_{x,a,b} p(b|x,a)
\end{equation}
with $p(b|x,a)$ being the conditional probability that Bob obtains the outcome $b$ given the preparation $x,a$.
The amount of certified randomness is then quantified by the conditional min-entropy as,
\begin{equation}
 H_{\min}=-\log_{2}P_{guess}   
\end{equation}

It is straightforward to see that the optimal classical success probability $\mathcal{P}_C^{\max}=7/12$ is attained by strategies in which Eve can perfectly predict some of Bob's output (see Tab. \ref{Tab:21_vertices}). This implies, $P_{\mathrm{guess}}=1$ and $H_{\min}=0$.

On the other hand, for the optimal quantum strategy derived in Appendix~\ref{Appx:qo} certifies the state and measurements as
\begin{eqnarray}
    \rho_{xa}=\frac{\mathbb{I}+(-1)^aA_{x \oplus_3 2a }}{2} ; \ \ E_b= \frac{\mathbb{I}+B_b}{3}\ \ \forall x,b\in\{0,1,2\}
\end{eqnarray}
where, $a\in{0,1}$ and $B_b=(A_b-A_{b\oplus_31})/\sqrt{3}$.

Therefore, the maximum guessing probability of Eve is given by 
\begin{eqnarray}
    P_{guess}=\max_{x,a,b} \Tr [\frac{\mathbb{I}+(-1)^aA_{x \oplus_3 2a}B_b}{6}]
\end{eqnarray}
The maximum value of $\Tr[A_{x \oplus_3 2a} B_b] = \Tr[A_{x \oplus_3 2a}(A_b-A_{b\oplus_31})]/\sqrt{3}$ is given by $\sqrt{3}$. Consequently, Eve's guessing probability becomes,

\begin{eqnarray}
    P_{guess}=\frac{2+\sqrt3}{6}
\end{eqnarray}

Hence, from the optimal value of the witness Eq.~\eqref{Avg_Success}, we can certify minimum $H_{min}=-\log_2P_{guess}\approx 0.69$ bit of randomness. It is worth emphasizing that the resulting quantum randomness is obtained without any seed randomness in a simple and experimentally friendly prepare-and-measure scenario.
}
\begin{comment}
\begin{figure}[H]
\centering
\begin{tikzpicture}
\begin{axis}[
  width=\columnwidth,
  height=0.62\columnwidth,
  xmin=0.86, xmax=1,
  ymin=0.60, ymax=0.85,
  domain=0.86:1,
  samples=300,
  axis lines=left,
  xlabel={$v$},
  ylabel={$-\log_2\!\bigl(\mathcal{P}_Q(v)\bigr)\;\text{(bits)}$},
  xtick={0.86,0.90,0.94,0.98,1.00},
  ytick distance=0.05,
  tick align=outside,
  tick style={black!70},
  label style={font=\small},
  grid=both,
  grid style={line width=.1pt, draw=gray!18},
  major grid style={line width=.2pt, draw=gray!35},
]
\addplot[blue!70!black, semithick, smooth]
  ({x},{-ln(x*(1/3*(1+sqrt(3)/2)) + (1-x)/3) / ln(2)});
\end{axis}
\end{tikzpicture}
\caption{Binary surprisal $-\log_2\bigl(\mathcal{P}_Q(v)\bigr)$ for $v\in[\frac{\sqrt{3}}{2},1]$, where $\mathcal{P}_Q(v)$ is defined in Eq.~\eqref{eq:vis} and $\mathcal{P}_Q^{\max}=\tfrac{1}{3}\left(1+\tfrac{\sqrt{3}}{2}\right)$.}
\label{fig:minuslogPQ}
\end{figure}
\end{comment}

\section{Coherence as the non-classical feature of measurements}

We emphasize that the communication advantage of a qubit achieved here does not come from  measurement incompatibility, as Bob uses only a single measurement described by a three-outcome qubit POVM.  {We note here that nonclassicalities in terms of contextuality \cite{Selby2023} and violation of macro-realism \cite{SSarkarr2025APLQ}  have been demonstrated using single measurement, by using  five and three outcome POVM respectively.} We show that such POVMs possess a distinctive ability to capture coherence, which, in our communication scenario, can be regarded as a form of nonclassicality. To demonstrate this, we invoke the resource theory of POVM-based coherence.

The resource theory of POVM-based coherence \cite{Bischof2019_PRL_Coherence,Bischof2021PhysRevA,SSarkar2025PRA_Qresources}, defined via their ability to detect coherence of quantum states, is inspired by the resource theory of state-based coherence \cite{aberg2006quantifyingsuperposition,plenioprl19} . Consider a POVM $E=\{E_{i}\}$ on $\mathcal{H}_{d}$ with $E_{i}\geq 0$, $\sum_{i}E_{i}=\mathbb{I}$ with $\mathcal{H}_{d}$ being the Hilbert space of dimension $d$. Now, according to Naimark extension any such POVM can be extended to a projective measurement $\textbf{P}=\{\pi_{i}\}$ on Hilbert space $\mathcal{H}_{d'}$, with $d'\geq d$. In this each of the projectors is written in terms of the basis vectors $\{\ket{\zeta_{i}}\}$ as $\pi_{i}= \ketbra{\zeta_{i}}$. Any state that is a convex mixture of those basis states is called incoherent states and is represented as,

\begin{eqnarray}
    \sigma_{inc}=\sum_{i}p_{i} \ketbra{\zeta_{i}}{\zeta_{i}}
\end{eqnarray}
 where $\sum_{i}p_{i}=1$. The incoherent states are produced by completely dephasing operation as $\sigma_{inc}= \Lambda (\sigma)= \sum_{i} \ketbra{\zeta_{i}}{\zeta_{i}}\sigma\ketbra{\zeta_{i}}{\zeta_{i}}$, where $\Lambda$ is a coherence destroying map.% In general, the orthogonal projectors $\pi_{i}$ can be of higher rank and in that case the incoherent states would be   diagonal instead diagonal \cite{aberg2006quantifyingsuperposition}. In the generalised resource theory of coherence, these   diagonal states serve as free objects, and  -coherence non-increasing operations constitute the free operations.

The largest class of free operations that cannot increase coherence, known as maximally block-incoherent operations represented by $\Lambda_{MBI}$. If $\mathcal{I}$ denotes the set of incoherent states, then $\Lambda_{MBI}[\mathcal{I}] \subseteq \mathcal{I}$.
 The standard measure of coherence that contained in a state $\sigma$ can be defined in terms of relative entropy\cite{aberg2006quantifyingsuperposition} as 
 \begin{eqnarray}
     C_{rel}(\sigma,\textbf{P})=S(\Lambda[\sigma])-S(\sigma) \label{coherence_PVM}
 \end{eqnarray}
Here, $S(\sigma)$ is the von Neumann entropy. This quantifier fulfills all the criteria required for a valid coherence measure \cite{Bischof2019_PRL_Coherence,Bischof2021PhysRevA}:  
i) Faithfulness: $C_{rel}(\sigma,\mathbf{P})>0$ and it becomes $0$ only when $\sigma=\sigma_{inc}$.  
ii) Monotonicity: $C_{rel}(\Lambda_{MBI} [\sigma],\mathbf{P})\leq C_{rel}(\sigma,\mathbf{P})$ and iii) Convexity: $C_{rel}\!\left(\sum_i p_i\sigma_i,\mathbf{P}\right)\leq \sum_i p_i C_{rel}(\sigma_i,\mathbf{P})\ \ \forall \{\sigma_i\}$, where $p_i$ are probabilities satisfying $\sum_i p_i=1$.

The state-based coherence measure is extended to the resource theory of the POVM-based coherence measure \cite{Bischof2019_PRL_Coherence,Bischof2021PhysRevA}. Such a measure is related to its projector based counterpart in Eq.~\eqref{coherence_PVM}  through,
\begin{equation}
 C(\sigma,E)=C(\mathcal{E}(\sigma),\textbf{P}),   
\end{equation}
 with $\mathcal{E}(\sigma)$ being the embedding of the state $\sigma$ to $\mathcal{H}_{d'}$ such that $\Tr[\sigma \ E_{i}]= \Tr[\pi_{i}\mathcal{E}(\sigma)]$ for all $E_{i} \in E$. The POVM-based coherence is defined in an extended Naimark space where $\pi_{i}$ is the projector on $\mathcal{H}_{d'}$ corresponding to $E_{i}$ and can be expressed as,
\begin{eqnarray}\label{eq:cohpovm}
     C_{rel}(\sigma,\textbf{E})= H(\{p_i\})+\sum_i p_iS(\sigma_i)-S(\sigma)\label{POVM_Coherence}
\end{eqnarray}
 where  $p_i=\Tr[\sigma E_i]$ is the probability of getting the outcome $i$, $\sigma_i$ is the corresponding post measurement state, and $H(\{p_i\})=-\sum_i p_i \log_2 p_i$ is the Shannon entropy. Crucially, the measure $ C_{rel}(\sigma,\textbf{E})$ in Eq.~\eqref{POVM_Coherence} is independent of the choice of Naimark extension \cite{Bischof2019_PRL_Coherence}.

Equipped with the above tools, we now demonstrate how coherence enables the quantum  advantage in the HFW scenario. We recall that in quantum theory, Alice prepares six qubit states $\{\rho_{xa}\}$ satisfy Eqs.~\eqref{Q_PO_Condition1} and \eqref{Q_PO_Condition2}. Bob implements the three-outcome qubit measurement $\mathbb{M}\equiv\{E_b | \ E_b=\alpha_{b} \pi_{b}, \ \sum_{b=0}^{2}E_{b}= \mathds{I} \}$ where $\alpha_b\in [0,1]$. We proved in Appendix~\ref{Appx:qo} that the maximum success probability of the communication game $\mathcal{P}^{max}_{Q}=\frac{1}{3}\left(1+ \frac{\sqrt{3}}{2} \right)\approx 0.62$ is attained when $\alpha_b=\frac{2}{3}\ \forall b\in\{0,1,2\} $,$i.e.$, the trine-spin POVM. Interestingly, for the extreme cases when one of the $\alpha_b = 0$, the quantum and classical success probabilities are the same; hence, no quantum advantage arises, as shown in Fig.~\ref{Q-C}. 

From Eq.~\eqref{POVM_Coherence}, it is evident that the value of the coherence measure is explicitly state-dependent. Consequently, distinct choices of $\sigma$ generally yield different coherence characteristics.  For our purpose, we restrict our attention to the maximally mixed qubit state $\frac{\mathbb{I}}{2}$.  Since the average state corresponding to optimal quantum strategy $\mathcal{P}^{max}_{Q}$ is $\frac{\mathbb{I}}{2}$, we adopt this state as reference and we evaluate $C_{rel}\left(\frac{\mathbb{I}}{2},\mathbb{M}\right)$.  We demonstrate how coherence serves as a nonclassical resource in such constrained HFW scenario proved in the following lemma,

\begin{Lemma}
     \label{Lemma_Coherence_as_Resource}
     A necessary condition for achieving quantum advantage $ \mathcal{P}_{Q}>  \mathcal{P}_{C}^{max}$  is that the POVM- based coherence measure $C_{rel}\left(\frac{\mathbb{I}}{2},\mathbb{M}\right)$ is nonzero, and the maximum quantum value $ \mathcal{P}_{Q}^{max}$ of the witness  is achieved when $C_{rel}\left(\frac{\mathbb{I}}{2},\mathbb{M}\right)$ reaches its maximum value.
 \end{Lemma}
\begin{proof}

    We  first evaluate $C_{rel}\left(\frac{\mathbb{I}}{2},\mathbb{M}\right)$ in  Eq.~\eqref{eq:cohpovm} for the maximally mixed state $\frac{\mathbb{I}}{2}$ for the measurement $\mathbb{M}$. For $\frac{\mathbb{I}}{2}$, the probability of obtaining a outcome $b$ is  $p_{b}=\frac{\alpha_b}{2}$. Moreover, since this POVM is extremal, the post-measurement state $\rho_{b}$ will be pure, and therefore we have $\forall b \ \ S(\rho_b)=0$. Then, from Eq.~\eqref{eq:cohpovm} the POVM-based coherence for the measurement $\mathbb{M}$ is given by,

\begin{eqnarray}
    C_{rel}\left(\frac{\mathbb{I}}{2},\mathbb{M}\right) =-\sum_b \frac{\alpha_b}{2} \log_2 \frac{\alpha_b}{2}-1
\end{eqnarray}
where, $\sum_{b}\alpha_{b}=2$ and $0 \leq \alpha_{b} \leq 1$. It is straightforward to verify that $C_{rel}\!\left(\frac{\mathbb{I}}{2},\mathbb{M}\right) = 0$ if one of the $\alpha_b$ is $0$, that is when the measurement reduces to a standard projective measurement with two possible outcomes. For such measurements that are unable to reveal the coherence of maximally mixed state, our constructed game does not exhibit any quantum advantage, as also depicted in Fig.~\ref{Q-C}. %In the limiting cases $\alpha_{b} = 0$ or $\alpha_{b} = 1$, the corresponding measurement reduces to a standard projective measurement with two possible outcomes. In this regime, the relative coherence evaluates to $C_{rel}\!\left(\frac{\mathbb{I}}{2},\mathbb{M}\right) = 0$, and, accordingly, our constructed game does not exhibit any quantum advantage in these cases.

We determine the point at which $C_{rel}\left(\frac{\mathbb{I}}{2},\mathbb{M}\right)$ is maximised. By using $\alpha_2 = 2 - \alpha_1 - \alpha_0$ and setting the first derivative to zero, $\frac{\partial C_{rel}\!\left(\frac{\mathbb{I}}{2},\mathbb{M}\right)}{\partial \alpha_b} = 0$  yields 
\begin{eqnarray}
    \ln{\frac{\alpha_b}{2}}=\ln{\frac{2-\alpha_0-\alpha_1}{2}}\ \ \forall b\in\{0,1\} \label{1st_derivative}
\end{eqnarray}
The Eq.~\eqref{1st_derivative} implies $2\alpha_0=2-\alpha_1$ and $2\alpha_1=2-\alpha_0$. This leads to the unique solution $\alpha_b=\frac{2}{3}$ $\forall b$, meaning  that the measurement $\mathbb{M}$ corresponds  the trine-POVM. We already proved that achieving the maximum quantum value $ \mathcal{P}_{Q}^{max}$ of the witness requires the measurement to be trine-POVM. 

\end{proof}

  Lemma~\ref{Lemma_Coherence_as_Resource} can be intuitively visualized by comparing Fig.~\ref{Q-C} and Fig.~\ref{Coherence}. These figures show that, in the absence of any quantum advantage, $C_{rel}\left(\frac{\mathbb{I}}{2},\mathbb{M}\right)$ vanishes, whereas whenever a quantum advantage is present, $C_{rel}\left(\frac{\mathbb{I}}{2},\mathbb{M}\right)$ assumes nonzero values. Moreover, when the quantum advantage is maximised, $C_{rel}\left(\frac{\mathbb{I}}{2},\mathbb{M}\right)$ also attains its maximum. Therefore, we argue that in scenarios without measurement incompatibility, the POVM based   coherence constitutes the relevant form of non-classicality responsible for the observed quantum advantage.
 %This is sufficient for the proof of Lemma~\ref{Lemma_Coherence_as_Resource}.
% In order to examine whether $E_{b}$ can be written in the form of Eq.~\eqref{eq:incoherent} we fix a particular basis as $\{\ket{\psi_{0}},\ket{\psi_{0}^{\perp}}\}$. It is straightforward to see that only at the extreme cases of $\alpha_{0}=0$  and $\alpha_{0}=1$, the POVM $E_{b}$ can be written in the form of Eq.\eqref{eq:incoherent}. However, these two extreme cases are trivial as they represent the two-outcome sharp qubit measurements. For any other values of $\alpha_b$, the measurement remains a three-outcome qubit measurement and is not a free POVM. Therefore, we argue that in the absence of measurement incompatibility, the ability of our three-outcome qubit POVM to detect the state coherence, is the non-classicality  that provides the quantum advantage.

\section{Summary and outlook}
We put forward a seemingly impossible demonstration of quantum supremacy of a qubit over a c-bit in the HFW scenario when SR is a free resource, but no additional quantum resources, such as entanglement or measurement incompatibility, are injected. The communication game introduced in this work  naturally fits within the HFW framework \cite{Holevo1973, FW_2015} augmented by cryptographic constraints on Alice's preparations, which we termed parity concealment constraints given in Eqs. (\ref{Parity_Constraint}) and (\ref{Parity_Constraint2}).  Imposing that constraint affects classical and quantum strategies in markedly different ways, ultimately exposing the communication supremacy of a qubit in the HFW scenario.  More precisely, we introduced a suitable witness—the success probability $( \mathcal{P})$ of the game, which enabled us to reveal qubit supremacy. We demonstrated that, for non-zero values of $\alpha_b$, we have $ \mathcal{P}_{Q}>  \mathcal{P}_{C}^{max}\approx 0.58$, and for $\alpha_b=2/3$ $\forall b$, the optimal quantum value  $ \mathcal{P}_{Q}^{max}=0.62$ is achieved. {Further, we demonstrate that the maximum quantum advantage can be exploited to certify $0.69$ bit of randomness. Importantly, in the HFW framework Bob has no input, thereby our protocol generates certified randomness  without seed randomness.}

We emphasize that the HFW scenario is, by construction, devoid of quantum resources such as entanglement or measurement incompatibility. This invites a deeper inquiry into what  aspects of non-classicality could be responsible for revealing quantum advantage here. We have demonstrated  that Bob's three-outcome qubit POVM possesses a special property of detecting coherence \cite{plenioprl19}, which can be regarded as the crucial nonclassicality needed to demonstrate qubit supremacy in the HFW scenario. Specifically, we showed that whenever $ \mathcal{P}_{Q}>  \mathcal{P}_{C}^{max}$ is obtained, Bob's measurement $\mathbb{M}$ detects the coherence; that is,  $C_{rel}\left(\frac{\mathbb{I}}{2},\mathbb{M}\right)$ is non-zero. For the case when one of $\alpha_{b} = 0$, the measurement $\mathbb{M}$ becomes projective, and the coherence measure $C_{rel}\left(\frac{\mathbb{I}}{2},\mathbb{M}\right)=0$. In that case, we found no quantum advantage, i.e., $ \mathcal{P}_{Q}=  \mathcal{P}_{C}^{max}$. 

Finally, we conjecture that for any fixed measurement empowered with state coherence detection capability, suitable constraints in preparation exist such that quantum supremacy in the HFW scenario can be demonstrated. Moreover, apart from the parity concealment constraint, a rich family of cryptographic constraints remains to be identified that can also establish qubit supremacy in the HFW scenario. Investigations in those directions could be an exciting avenue for future research.
%More broadly we anticipate that apart form merely such {parity concealment constraints}, a rich family of cryptographic constraints remains to be identified that can reestablished qubit supremacy in HFW scenario, could be an exciting direction for future research.

%Furthermore, we emphasize that the HFW scenario is, by construction devoid of other common quantum resources such as entanglement and incompatibility. This  absence invites a deeper inquiry into the essential aspects of non-classicality required to manifest quantum advantage in this context. We have demonstrated  that POVM based block coherence \cite{plenioprl19} is the crucial resource to demonstrate qubit advantage in HFW scenario. We also conjecture that for any fixed measurement empowered with state coherence detection capability, suitable constraints in preparation exist such that quantum supremacy in the HFW scenario can be demonstrated. Investigations following this path could be an exciting direction for future research.

    {\appendix
     \section{Proof of Lemma 1(i): Derivation of the maximum success probability using c-bit communication aided  by SR}\label{optimisation of bit comm}
Let us first analyse the most general strategy when Alice sends one c-bit to Bob without using any SR. Assume that before starting the game, Alice and Bob agree upon an encoding ($\mathds{E}_1$) and decoding ($\mathds{D}_1$) strategy as given below.\\

\textbf{Encoding strategy ($\mathds{E}_1$)}: For the input $xa$, Alice prepares '$0$' with probability $p_{xa}$ and '$1$' with probability $1-p_{xa}$.\\

 \textbf{Decoding strategy ($\mathds{D}_1$)}: If Bob receives '$0$' he outputs $0$ with probability $s_0$, $1$ with probability $s_1$ and $2$ with probability $s_2$. Similarly, if he receives '$1$', he outputs $b$ with probability $r_b, \ \forall b\in \{0,1,2\}$ with $\sum_b s_b=\sum_b r_b=1$. An explicit description of the strategy is given in the following table.\\

\begin{center}
\resizebox{\columnwidth}{!}{
\begin{tabular}{||c| c c c c c c| c ||} 
 \hline
 Message & 00 & 01 & 10 & 11 & 20 & 21 & Decoding \\ [0.5ex] 
 \hline\hline
 0 & $p_{00}$ & $p_{01}$ & $p_{10}$ &$p_{11}$ & $p_{20}$ & $p_{21}$ & $s_0\rightarrow0$;\ $s_1 \rightarrow 1$;\ $s_2\rightarrow 2$ \\ 
 \hline
 1 & 1-$p_{00}$ & 1-$p_{01}$ & 1-$p_{10}$ & 1-$p_{11}$ & 1-$p_{20}$ & 1-$p_{21}$ & $r_0\rightarrow0$;\ $r_1 \rightarrow 1$;\ $r_2\rightarrow 2$ \\[1ex] 
 \hline 
\end{tabular}}
\end{center}
In order to satisfy the restriction Eq.(\ref{Parity_Constraint}) and Eq.(\ref{Parity_Constraint2}) of the game, Alice's preparation adhere,
 \begin{eqnarray}\label{c1_parity}
&&\frac{1}{3}\left(p_{00}+p_{10}+p_{20}\right)= \frac{1}{3}\left(p_{01}+p_{11}+p_{21}\right)\equiv\kappa_1\\
\label{c2_parity}
&&\frac{1}{2}\left(p_{00}+p_{11}\right)=\frac{1}{2}\left(p_{01}+p_{20}\right)=\frac{1}{2}\left(p_{10}+p_{21}\right)\equiv\kappa_2
\end{eqnarray}
where $\kappa_1, \kappa_2\in [0,1]$. It is simple to check from Eqs. (\ref{c1_parity}) and (\ref{c2_parity}) that $\kappa_1=\kappa_2$.

Now, consider that Alice and Bob share an unbounded amount of SR denoted by the variable $\tau$ with probability $p(\tau)$ and use this correlation to boost the winning possibility of the game. A general approach could be to fix an encoding strategy ($\mathds{E}_{\tau}$) and decoding strategy ($\mathds{D}_{\tau}$)  corresponding to each $\tau$. {Note that $\tau$ is common knowledge to both the parties Alice and Bob. Therefore, in any given run, an encoding ($\mathds{E}_{\tau}$) that leaks parity information, would immediately compromise the parity obliviousness of inputs because Bob also have direct access to $\tau$. Hence, to keep her privacy, Alice must choose such encoding that satisfy Eqs. (\ref{c1_parity}) and (\ref{c2_parity}) for all $\tau$.}  The average success probability ($\mathcal{P}_C$) of the game due to Alice's one $c$-bit communication aided with SR  becomes,
\begin{eqnarray}\label{c3}
\mathcal{P}_C&&=\sum_{\tau}p(\tau)\mathcal{P}(\mathds{E}_{\tau},\mathds{D}_{\tau})\nonumber \\
    &&=\frac{1}{6}\sum_{\tau}p(\tau)([p_{00}(\tau)+p_{21}(\tau)]s_0(\tau) +[p_{11}(\tau)+p_{20}(\tau)]s_2(\tau)\nonumber\\
     &&+ [p_{01}(\tau)+p_{10}(\tau)]s_1(\tau)+  [2-p_{00}(\tau)-p_{21}(\tau)]r_0(\tau) \nonumber\\
     &&+[2-p_{11}(\tau)-p_{20}(\tau)]r_2(\tau)+ [2-p_{01}(\tau)-p_{10}(\tau)]r_1(\tau))\nonumber\\
\end{eqnarray}
where $p_{xa}(\tau)$ is the short-hand notation for the probability $p(0|xa,\tau)$ corresponding to Alice's message $'0'$ for the input $(x,a)$ and the shared variable $\tau$ and similarly $s_{b}(\tau)$ and $r_{b}(\tau)$ are the probabilities of Bob's output $'0'$ and $'1'$ respectively given $\tau$. Using Eq.\eqref{c2_parity} we eliminate $p_{01}(\tau),p_{11}(\tau),p_{21}(\tau)$ and substitute $s_2(\tau)=1-s_1(\tau)-s_0(\tau)$, $r_2(\tau)=1-r_1(\tau)-r_0(\tau)$ in Eq.\eqref{c3}, we finally obtain,
 \begin{eqnarray}
     \mathcal{P}_C=&&\frac{1}{3} + \frac{1}{6} \sum_{\tau}p(\tau)\Big[\big(2p_{00}(\tau)-p_{10}(\tau)-p_{20}(\tau)\big)\big(s_0(\tau)-r_0(\tau)\big) \nonumber\\
     &&+\big(p_{10}(\tau)+p_{00}(\tau)-2p_{20}(\tau)\big)\big(s_1(\tau)-r_1(\tau)\big)\Big]
 \end{eqnarray}
Subsequently, using Eq.\eqref{c1_parity} we get,
 \begin{eqnarray}
 \label{cc}
     \mathcal{P}_C&&=\frac{1}{3}+ \frac{1}{2}\sum_{\tau} p(\tau)\Big[\big(p_{00}(\tau)-\kappa_2(\tau) \big)\big(s_0(\tau)-r_0(\tau)\big) \nonumber\\
     &&+  \big(\kappa_2(\tau)-p_{20}(\tau)\big)\big(s_1(\tau)-r_1(\tau)\big)\Big]
 \end{eqnarray}
 From Eq.~(\ref{cc}), it can be seen that there are four possibilities of values, (I)  $s_0(\tau)\geq r_0(\tau),   s_1(\tau) \geq r_1(\tau) $, (II) $s_0(\tau) \geq r_0(\tau),   s_1(\tau)\leq r_1(\tau) $, (III) $s_0(\tau)\leq r_0(\tau),   s_1(\tau) \geq r_1(\tau) $, and (IV) $s_0(\tau)\leq r_0(\tau),   s_1(\tau)\leq r_1(\tau) $. Let us now analyse the maximum values for all four cases.

For case (I), maximisation requires $r_0(\tau)=r_1(\tau)= 0$ and $s_0(\tau)= 1$, leading to $s_1(\tau)=0$. Eq.~(\ref{cc}) then reduces to 
\begin{eqnarray}
 \label{cc1}
     \mathcal{P}^{(I)}_c=\frac{1}{3}+ \frac{1}{2}\sum_{\tau\in (I)} p(\tau)\Big[p_{00}(\tau)-\kappa_2(\tau) \Big]
 \end{eqnarray}
For maximisation, we further take $p_{00}(\tau)=1$ and following Eq.\eqref{c2_parity} the minimum value of $\kappa_2(\tau)$ can be $\half$. Putting these together, we get 
\begin{eqnarray}
    \mathcal{P}_C^{(I)}=\frac{1}{3}+ \frac{1}{2}\sum_{\tau \in (I)}\half p(\tau) 
\end{eqnarray}
Similarly, for other three cases it is straightforward to check that the result remains the same as above $i.e.$, $\mathcal{P}_C^{(II)}=\frac{1}{3}+ \frac{1}{2}\sum\limits_{\tau \in (II)}\half p(\tau)$, $\mathcal{P}_C^{(III)}=\frac{1}{3}+ \frac{1}{2}\sum\limits_{\tau \in (III)}\half p(\tau)$ and $\mathcal{P}_C^{(IV)}=\frac{1}{3}+ \frac{1}{2}\sum\limits_{\tau \in (IV)}\half p(\tau)$. Considering all those relations, we find the maximum value, $(\mathcal{P}_C)^{max}$ of $\mathcal{P}_C$ to be 
\begin{eqnarray}
\left(\mathcal{P}_C\right)^{max}=\frac{1}{3} + \frac{1}{4} \sum_{\tau} p(\tau) =\frac{7}{12}\approx 0.58
\end{eqnarray} 
Therefore, the maximum success probability achieved for the communication game for Alice's one $c$-bit communication assisted with unbounded SR is $0.58$. % \textcolor{red}{Then the maximum success probability (as in Lemma. \textbf{1} of the main text) of the communication game using qubit communication surpasses the one achieved with $c$-bit, establishing the quantum supremacy in HFW scenario.}

%\section{Where we stand with respect to Holevo-Frankle-Weiner results}

{
\section{Full characterization of classical correlation}\label{Ap:Polytope}

In order to characterize the classical correlations arising in the communication game, we need to determine the geometry of the set of probabilities  $\{p(b|x,a)\}$ , where, 

\begin{eqnarray}
p(b\mid x,a)=\sum_{m\in\{0,1\}} p(m\mid x,a)\,p(b\mid m).
\label{eq:cC}
\end{eqnarray}

Let us first denote the set of probabilities consisting of $p(0\mid x,a)=p_{xa}$ and $p(1\mid x,a)=1-p_{xa}$, associated with the encoding part. These probabilities respect the parity concealment constraints as given by Eqs.~\eqref{c1_parity} and ~\eqref{c2_parity}. Rewriting these constraints explicitly we get,

\begin{subequations}
\begin{align}
 p_{00}-p_{01}+p_{10}-p_{11}+p_{20}-p_{21} &= 0,\\
 p_{00}-p_{01}+p_{11}-p_{20} &= 0,\\
 p_{01}-p_{10}+p_{20}-p_{21} &= 0.
\end{align}
\end{subequations}

Among these six probabilities, only three are independent, while the remaining three can be expressed as linear combinations of the former three, namely,
\begin{subequations}\label{Eq:C3}
    \begin{eqnarray}
    p_{21}&=& 2p_{00}-2p_{01}+p_{10}\\
    p_{20} &=& 2p_{00}-3p_{01}+2p_{10} \\
    p_{11} &= &p_{00}-2p_{01}+2p_{10}
    \end{eqnarray}
\end{subequations}

Then this set of probabilities $\{p_{xa}\}$ forms a polytope in dimension three and can be defined formally as follows,

\begin{equation}
\mathcal{E}_{\mathrm C}
=
\left\{
(p_{00},p_{01},p_{10})\in\mathbb{R}^3 :
\begin{array}{l}
0\le p_{00}\le1,\\
0\le p_{01}\le1,\\
0\le p_{10}\le1,\\
0\le p_{00}-2p_{01}+2p_{10}\le1,\\
0\le 2p_{00}-3p_{01}+2p_{10}\le1,\\
0\le 2p_{00}-2p_{01}+p_{10}\le1
\end{array}
\right\}.
\end{equation}

The vertices of this encoding polytope $(p_{00},p_{01},p_{10})\in\mathbb{R}^{3}$ are given by,

\begin{align}
v_1 &= \left(0,0,0\right), \ \
v_2 = \left(0,0,\frac{1}{2}\right), \ \
v_3 = \left(0,\frac{1}{2},1\right), \nonumber\\
v_4 &= \left(\frac{1}{2},0,0\right), \ \  
v_5 = \left(\frac{1}{2},1,1\right), \ \ 
v_6 = \left(1,\frac{1}{2},0\right), \nonumber\\ 
v_7 &= \left(1,1,\frac{1}{2}\right), \ \ 
v_8 = \left(1,1,1\right).
\end{align}

Any probability distribution corresponding to Alice's encoding strategies must be inside this polytope.

It is important to note that, in the absence of the parity concealment constraint, the classical encoding polytope is the six-dimensional hypercube, whose vertices correspond to deterministic encodings. Imposing the constraints amounts to intersecting this hypercube with an affine subspace representing the constraints. Consequently, the vertices of the constrained polytope are no longer deterministic encodings; six of the eight vertices are intrinsically stochastic, containing probabilities equal to 1/2.  Therefore, the nontrivial vertices of the encoding polytope cannot be expressed as convex combinations of admissible deterministic encodings.

Now, consider Bob's decoding function  $D=\{D(0),D(1)\}$, where $D(m)\in \{0,1,2\}$. Then for each input from the set  $\{xa\}$ we have,
\begin{eqnarray}\label{Eq: decoding_poly}
    p(b \mid x,a) &=& p(m=0 \mid x,a)  \delta_{b, D(0)} + p(m=1 \mid x,a)  \delta_{b, D(1)} \nonumber \\
    & \equiv& p_{xa}  (\delta\label{Eq:C5}_{b, D(0)}-\delta_{b, D(1)})+ \delta_{b, D(1)}\label{Eq:C5}
\end{eqnarray}

Note that, a total of $18$ variables needed to represent the space of probabilities formed by $\{p(b\mid x,a)\}$. However, not all the variables are independent. Normalization, $\sum_{b=0,1,2} p(b\mid x,a)=1$, reduce $6$ variables. By using Eq.~\eqref{Eq:C3} we rewrite  Eq.~\eqref{Eq:C5} as,
{\small \begin{subequations}\label{Eq:C6}
    \begin{eqnarray}
    p(b\mid 2,1)&=& 2p(b\mid 0,0)-2p(b\mid 0,1)+p(b\mid 1,0)\\
    p(b\mid 2,0) &=& 2p(b\mid 0,0)-3p(b\mid 0,1)+2p(b\mid 1,0) \\
    p(b\mid 1,1) &= &p(b\mid 0,0)-2p(b\mid 0,1)+2p(b\mid 1,0)
    \end{eqnarray}
\end{subequations}}
where $b\in \{0,1,2\}$.
\begingroup
\begin{table}[t]
\centering
\caption{All $21$ vertices of the classical polytope and corresponding success probabilities.}
\label{Tab:21_vertices}
\normalsize
\setlength{\tabcolsep}{4pt}
\renewcommand{\arraystretch}{1.15}
\begin{tabular}{*{7}{c}}
\toprule
$\mathcal{P}_{C}$ & $p(0|00)$ & $p(1|00)$ & $p(0|01)$ & $p(1|01)$ & $p(0|10)$ & $p(1|10)$ \\
\midrule
\multirow{6}{*}{$7/12$}
 & $1$ & $0$ & $1$ & $0$ & $1/2$ & $0$ \\
 & $0$ & $1$ & $0$ & $1$ & $0$ & $1/2$ \\
 & $1$ & $0$ & $1/2$ & $1/2$ & $0$ & $1$ \\
 & $1$ & $0$ & $1/2$ & $0$ & $0$ & $0$ \\
 & $1/2$ & $1/2$ & $0$ & $1$ & $0$ & $1$ \\
 & $0$ & $1/2$ & $0$ & $1$ & $0$ & $1$ \\
\midrule
\multirow{9}{*}{$1/3$}
 & $1$ & $0$ & $1$ & $0$ & $1/2$ & $1/2$ \\
 & $0$ & $1$ & $0$ & $1$ & $1/2$ & $1/2$ \\
 & $0$ & $1$ & $0$ & $1/2$ & $0$ & $0$ \\
 & $0$ & $0$ & $0$ & $1/2$ & $0$ & $1$ \\
 & $1/2$ & $0$ & $1$ & $0$ & $1$ & $0$ \\
 & $1/2$ & $0$ & $0$ & $0$ & $0$ & $0$ \\
 & $1$ & $0$ & $1$ & $0$ & $1$ & $0$ \\
 & $0$ & $1$ & $0$ & $1$ & $0$ & $1$ \\
 & $0$ & $0$ & $0$ & $0$ & $0$ & $0$ \\
\midrule
\multirow{6}{*}{$1/12$}
 & $0$ & $0$ & $0$ & $0$ & $1/2$ & $0$ \\
 & $0$ & $0$ & $0$ & $0$ & $0$ & $1/2$ \\
 & $0$ & $1$ & $1/2$ & $1/2$ & $1$ & $0$ \\
 & $0$ & $0$ & $1/2$ & $0$ & $1$ & $0$ \\
 & $1/2$ & $1/2$ & $1$ & $0$ & $1$ & $0$ \\
 & $0$ & $1/2$ & $0$ & $0$ & $0$ & $0$ \\
\bottomrule
\end{tabular}
\end{table}
\endgroup

These relations eliminate six additional dependent variables. Consequently, the classical correlation polytope is embedded in a six-dimensional probability space. Clearly, for a fixed encoding, there exist $3^2=9$ possible decoding strategies. Therefore, for the $8$ encoding strategies corresponding to the aforementioned vertices, a total of $8 \times 9= 72$ admissible strategies possible for the outcome correlation $p(b|x,a)$. After removing redundancy, we found $21$ strategies for $\{p(b|x,a)\}$ represented by the vertices $\{w_{i}\}$ of the polytope given by,
\begin{equation}
\mathcal{K}_{\mathrm{C}}
=\operatorname{Conv}
\left\{
\mathbf{w}_1,\mathbf{w}_2,\ldots,\mathbf{w}_{21}
\right\} \subset \mathbb{R}^{6},
\end{equation}
where each of the vertex has the vector form given by 
\begin{align}
\hspace{-0.25cm}w_{i}\hspace{-0.1cm}=\hspace{-0.1cm}\Big(p(0|00),p(1|00),p(0|01),p(1|01),p(0|10),p(1|10)\Big)
\end{align}
The entries of the vectors are explicitly listed in Table \ref{Tab:21_vertices}.

From each of the $21$ vertices we calculate the average success probability given by Eq.~\eqref{Avg_Success} and the maximum is obtained as $\mathcal{P}_C^{max}=\tfrac{7}{12}$. As the witness provided in Eq.~\eqref{Avg_Success} is linear, it should be maximized at least by one of the vertices. Hence, by utilizing shared randomness ($\tau$) , one can only obtain the probabilities within the polytope which are convex mixture of the probabilities corresponding to the vertices. Therefore, the maximum average success probability achievable in this scenario is  $\tfrac{7}{12}$, even if unbounded shared randomness is used.
\section{An example of classical encoding-decoding}\label{ced}
Here, we focus on the vertex $w_3=(1,0,\half,\half,0,1)$ from Table~\ref{Tab:21_vertices} and present the corresponding explicit strategy for better understanding. }
\begin{table}[H]
\centering
\setlength{\tabcolsep}{4pt} 
\renewcommand{\arraystretch}{1.2}  
\begin{tabular}{|c|c|c|c|c|}
 \hline 
\textbf{\shortstack{Alice's\\Input\\String ($xa$)}} &
\textbf{\shortstack{Alice's\\Message\\($m$)}} &
\textbf{\shortstack{Bob's\\Decoding\\($b$)}} &
\textbf{\shortstack{Success\\Condition\\$b = x \oplus_3 a$}} &
\textbf{\shortstack{Average\\Success\\$\mathcal{P}_C$}} \\ \hline
00 & 0   & 0   & \checkmark & \multirow{6}{*}{$\frac{7}{12} \approx 0.58$} \\ \cline{1-4}
01 & 0/1 & 0$\rightarrow$0,1$\rightarrow$1 & $\times$ / \checkmark & \\ \cline{1-4}
10 & 1   & 1   & \checkmark & \\ \cline{1-4}
11 & 1   & 1   & $\times$ & \\ \cline{1-4}
20 & 0/1 & 0$\rightarrow$0, 1$\rightarrow$1 & $\times$ & \\ \cline{1-4}
21 & 0   & 0   & \checkmark & \\ \hline 
\end{tabular}
\caption{An example of an encoding–decoding strategy achieving the optimal classical success probability.}
\label{tab:game}
\end{table}

\section{Detailed proof of Lemma  1(ii): Derivation of maximum quantum success probability  $\mathcal{P}_{Q}^{max}$}\label{Appx:qo}
Without loss of generality consider that Alice prepares the following six qubit states 
\ba\label{P1}
\rho_{x0}=\frac{\mathds{I}+ A_x}{2}; \ \ \  \rho_{x1}=\frac{\mathds{I} - A_{x \oplus_3 2}}{2}, \ \ x \in \{0,1,2\}
\ea
satisfying the restriction of Eq.~\eqref{Q_PO_Condition2} of the main text, so that $\half(\rho_{x0}+\rho_{x1})\equiv \rho_1, \ \forall x$.  To satisfy the  restriction of Eq.~\eqref{Q_PO_Condition1} of the main text, we need
\ba\label{P2}
\sum_{x=0,1,2}\frac{1}{3}\rho_{x0}=\sum_{x=0,1,2} \frac{1}{3}\rho_{x1}\equiv \rho_{2} 
\ea
which implies that
\ba\label{Cnstrn 2_Observable  form}
 A_0+A_1+A_2=0 
\ea
    
Bob chooses a general three-outcome qubit POVM as given by,
\ba
E_0=\alpha_0\frac{\mathds{I}+ B_0}{2};\ \ E_1= \alpha_1\frac{\mathds{I}+ B_1}{2};\ \ E_2=\alpha_2\frac{\mathds{I}+ B_2}{2} 
\ea
The completeness relation of measurement, $E_0+E_1+E_2=\mathds{I}$ implies
\begin{eqnarray}\label{Condition of Bob's Observables}
        \alpha_0B_0+\alpha_1B_1 +\alpha_2 B_2=0 , \ \ \ \text{and} \ \  \alpha_0+\alpha_1+\alpha_2=2
\end{eqnarray}
Now, using Eqs.\eqref{P1} and \eqref{P2} , the quantum success probability in Eq.~\eqref{Avg_Success} of the main text can be explicitly written as,

\ba\label{Avg_Succ_11}
\mathcal{P}_{Q}&=\frac{1}{6}\Tr[(\rho_{00}+\rho_{21})E_0+ (\rho_{10}+ \rho_{01})E_1+ (\rho_{20}+ \rho_{11})E_2] &\nonumber\\
& = \frac{1}{6}\Tr\big[\frac{\alpha_0}{4}(2\mathds{I}+ A_0- A_1)( \mathds{I}+B_0)+ \frac{\alpha_1}{4}(2\mathds{I}+ A_1- A_2)(\mathds{I}+B_1)&\nonumber\\ &+ \frac{\alpha_2}{4} (2\mathds{I}+ A_2- A_0)(\mathds{I}+B_2)\big]&.
\ea\label{Avg_Succ_12}
Here we define three normalisation constants $\omega_0=||A_0 - A_1||,\omega_1=||A_1 - A_2||,\omega_2=||A_2 - A_0||$, \Big($||\mathcal{N}||=\sqrt{\Tr (\mathcal{N}^{\dagger}\mathcal{N})}$ is Frobenius norm of $\mathcal{N}$\Big), such that  Eq.\eqref{Avg_Succ_11} can be rewritten as,
\ba
\label{pqq}
\resizebox{\columnwidth}{!}{\ensuremath{\mathcal{P}_{Q}= \frac{1}{3}+ \frac{1}{24}\Tr\bigg[\alpha_0\frac{A_0-A_1}{\omega_0}B_0\omega_0 +  \alpha_1 \frac{A_1-A_2}{\omega_1} B_1 \omega_1 +  \alpha_2 \frac{A_2- A_0}{\omega_2}B_2\omega_2\bigg].}}\nonumber\\
\ea

It is now evident from Eq.\eqref{pqq} that the maximisation of $\mathcal{P}_{Q}$ demands,
\ba
B_0=\frac{A_0 - A_1}{\omega_0}; \  B_1=\frac{A_1 - A_2}{\omega_1}; \ 
 B_2=\frac{A_2-A_0}{\omega_2}. \nonumber
\ea
Since $(B_{y})^{2}=\mathbb{I}$, we get
\ba\label{A8}
\mathcal{P}_{Q} &&\leq \frac{1}{3} + \frac{1}{12}\max[\alpha_0\omega_0 + \alpha_1 \omega_1 + \alpha_2 \omega_2]\\ \nonumber
&&= \frac{1}{3}+ \frac{1}{12}\max\Big[\alpha_0\sqrt{\langle 2\mathds{I} -\{A_0,A_1\}\rangle} +\alpha_1\sqrt{\langle 2\mathds{I} - \{A_1,A_2\}\rangle} \nonumber\\ &&+\alpha_2\sqrt{\langle 2\mathds{I} -\{A_2,A_0\} \rangle} \Big]
\ea

Again from Eq.\eqref{Cnstrn 2_Observable  form} we have,
\begin{eqnarray}\label{cnstrn 1}
    3\mathds{I}+\{A_0 ,A_1\}+\{A_1 ,A_2\}+\{A_2,A_0\}=0
\end{eqnarray} 
Multiply $A_0$ from the right of Eq.~\eqref{Cnstrn 2_Observable  form} followed by multiplying same $A_0$ from the left and then summing both equations we obtain, 
\begin{eqnarray} \label{cnstrn 2}
    2\mathds{I} + \{A_0 ,A_1\}+\{A_0 ,A_2\} =0
\end{eqnarray}
Comparing both the Eqs.~\eqref{cnstrn 1} -\eqref{cnstrn 2} we get $\langle\{A_1 ,A_2\}\rangle=-1$. Similarly, we can obtain $\langle\{A_0 ,A_1\}\rangle=-1$ and $\langle\{A_0 ,A_2\}\rangle=-1$, following  the constrain Eq.\eqref{Cnstrn 2_Observable  form}. Those relations indicate $\omega_0=\omega_1=\omega_2=1/\sqrt{3}$.  Substituting those in Eq.\eqref{A8} we obtain the bound for success probability as,
\begin{eqnarray}
    \mathcal{P}_{Q}\leq \frac{1}{3}\left(1+ \frac{\sqrt{3}}{2} \right)
\end{eqnarray}

Here, Bob's observables corresponding to the optimal quantum strategy is given by,
\begin{eqnarray}\label{Bob's Observables}
    B_0=\frac{1}{\sqrt{3}}(A_0-A_1);\ \ B_1=\frac{1}{\sqrt{3}}(A_1-A_2);\ \ B_2=\frac{1}{\sqrt{3}}(A_2-A_0) \nonumber\\
\end{eqnarray}

Replacing all $B_b$ from Eq.\eqref{Bob's Observables} into Eq.\eqref{Condition of Bob's Observables} and after rearranging we obtain,
\begin{eqnarray}
    (\alpha_0-\alpha_2)A_0+(\alpha_1-\alpha_0)A_1+ (\alpha_2-\alpha_1)A_2=0 \nonumber
\end{eqnarray}
Right away, substituting $A_2=-A_0-A_1$ and multiplying $A_0$ from the right side followed by multiplying the same $A_0$ from the left and then summing both equations we get,
\begin{eqnarray}\label{A12}
    2\mathds{I}(\alpha_0+\alpha_1-2\alpha_2)+\{A_0,A_1\}(2\alpha_1-\alpha_0-\alpha_2)=0 
    \end{eqnarray}
Since $\langle\{A_0 ,A_1\}\rangle=-1$, it follows from Eq.\eqref{A12} that
 $\alpha_0=\alpha_2$. Similarly, substituting $A_0=-A_2-A_1$ and multiplying $A_1$ from the right side followed by multiplying the same $A_1$ from the left and then summing both equations and then using the relation $\langle\{A_2,A_1\}\rangle=-1$ we find $\alpha_0=\alpha_1$. Consolidating these  relations,   we obtain $\alpha_0=\alpha_1=\alpha_2$. Again we have $\alpha_0+\alpha_1+\alpha_2=2$ that straightforwardly indicates $ \alpha_0=\alpha_1=\alpha_2=\frac{2}{3}$. Hence, a suitable choice of states and observables which satisfy these relations can give maximum success probability of the game as $\mathcal{P}^{max}_{Q}=\frac{1}{3}\left(1+ \frac{\sqrt{3}}{2} \right)$.

 One of the examples that satisfy the required properties for maximum success probability is given here: Suppose Alice chooses $A_x= \cos\theta_x \sigma_Z+ \sin\theta_x \sigma_X $ with $\theta_x=\frac{2\pi x}{3}$. 
 From Eq.\eqref{Bob's Observables} we obtain $B_b=\cos\theta_b \sigma_Z- \sin\theta_b \sigma_X $ where $\theta_b=\frac{2\pi}{3}(\frac{1}{4}+b)$. After sating $\alpha_0=\alpha_1=\alpha_2=\frac{2}{3}$ simple calculation of $\mathcal{P}_Q$ gives $\mathcal{P}^{max}_{Q}$.
 \section*{Data Availability}
The authors declare that all data supporting these findings are available within this article.
\section*{Author contributions}
P.P. , S.M. and A.K.P. conceived the project. P.P. and S.M. carried out the technical parts of the proofs. A.K.P. supervised the project. All authors extensively analysed the results and contributed to writing the manuscript.
\section*{Acknowledgments}
P.P. acknowledges funding from the University Grants Commission (NTA Ref. No.-231610049800), Govt. of India. S.M. acknowledges ANRF National Post-Doctoral Fellowship (NPDF) under File No. PDF/2025/000262 and the local hospitality from the grant IITH/SG160 of IIT Hyderabad, India. A.K.P. acknowledges the research grant I-HUB/PDF/2022-23/06, Government of India.

\bibliography{ref}

@article{Baek_2025,
doi = {10.1088/1367-2630/adcd5a},
url = {https://doi.org/10.1088/1367-2630/adcd5a},
year = {2025},
month = {may},
publisher = {IOP Publishing},
volume = {27},
number = {5},
pages = {053001},
author = {Baek, Kyunghyun and Ryu, Junghee and Lee, Jinhyoung},
title = {Robustness measures for quantifying nonlocality},
journal = {New Journal of Physics},
}

@article{vidal99,
  title = {Robustness of entanglement},
  author = {Vidal, Guifr\'e and Tarrach, Rolf},
  journal = {Phys. Rev. A},
  volume = {59},
  issue = {1},
  pages = {141--155},
  numpages = {0},
  year = {1999},
  month = {Jan},
  publisher = {American Physical Society},
  doi = {10.1103/PhysRevA.59.141},
  url = {https://link.aps.org/doi/10.1103/PhysRevA.59.141}
}

@article{Holevo1973,
  title = {Bounds for the Quantity of Information Transmitted by a Quantum Communication Channel},
  author = {A.~S.~Holevo},
  journal = {Problems Inform. Transmission},
  volume = {9},
  year = {1973},
  publisher = {American Physical Society},
  url = {http://mathscinet.ams.org/mathscinet-getitem?mr=456936}
}

@article{densed_coding_92,
  title = {Communication via one- and two-particle operators on Einstein-Podolsky-Rosen states},
  author = {Bennett, Charles H. and Wiesner, Stephen J.},
  journal = {Phys. Rev. Lett.},
  volume = {69},
  issue = {20},
  pages = {2881--2884},
  numpages = {0},
  year = {1992},
  month = {Nov},
  publisher = {American Physical Society},
  doi = {10.1103/PhysRevLett.69.2881},
  url = {https://link.aps.org/doi/10.1103/PhysRevLett.69.2881}
}

@article{FW_2015,
author={Frenkel, Péter E. and Weiner, Mihály},
title={Classical Information Storage in an n-Level Quantum System},
journal={Communications in Mathematical Physics},
year={2015},
month={Dec},
day={01},
volume={340},
number={2},
pages={563-574},
issn={1432-0916},
doi={10.1007/s00220-015-2463-0},
url={https://doi.org/10.1007/s00220-015-2463-0}
}

@article{Wiesner1983RAC,
author = {Wiesner, Stephen},
title = {Conjugate coding},
year = {1983},
issue_date = {Winter-Spring 1983},
publisher = {Association for Computing Machinery},
address = {New York, NY, USA},
volume = {15},
number = {1},
issn = {0163-5700},
url = {https://doi.org/10.1145/1008908.1008920},
doi = {10.1145/1008908.1008920},
journal = {SIGACT News},
month = jan,
pages = {78–88},
numpages = {11}
}

@inproceedings{Ambainis1999RAC,
author = {Ambainis, Andris and Nayak, Ashwin and Ta-Shma, Ammon and Vazirani, Umesh},
title = {Dense quantum coding and a lower bound for 1-way quantum automata},
year = {1999},
isbn = {1581130678},
publisher = {Association for Computing Machinery},
address = {New York, NY, USA},
url = {https://doi.org/10.1145/301250.301347},
doi = {10.1145/301250.301347},
booktitle = {Proceedings of the Thirty-First Annual ACM Symposium on Theory of Computing},
location = {Atlanta, Georgia, USA},
series = {STOC '99}
}

@article{Ambainis2002RAC,
author = {Ambainis, Andris and Nayak, Ashwin and Ta-Shma, Amnon and Vazirani, Umesh},
title = {Dense quantum coding and quantum finite automata},
year = {2002},
issue_date = {July 2002},
publisher = {Association for Computing Machinery},
address = {New York, NY, USA},
volume = {49},
number = {4},
issn = {0004-5411},
url = {https://doi.org/10.1145/581771.581773},
doi = {10.1145/581771.581773},
journal = {J. ACM},
month = jul,
pages = {496–511},
numpages = {16}
}

@article{Selby2023,
  title = {Contextuality without Incompatibility},
  author = {Selby, John H. and Schmid, David and Wolfe, Elie and Sainz, Ana Bel\'en and Kunjwal, Ravi and Spekkens, Robert W.},
  journal = {Phys. Rev. Lett.},
  volume = {130},
  issue = {23},
  pages = {230201},
  numpages = {7},
  year = {2023},
  month = {Jun},
  publisher = {American Physical Society},
  doi = {10.1103/PhysRevLett.130.230201},
  url = {https://link.aps.org/doi/10.1103/PhysRevLett.130.230201}
}

@article{Carmeli_2020,
doi = {10.1209/0295-5075/130/50001},
url = {https://dx.doi.org/10.1209/0295-5075/130/50001},
year = {2020},
month = {jun},
publisher = {EDP Sciences, IOP Publishing and Società Italiana di Fisica},
volume = {130},
number = {5},
pages = {50001},
author = {Claudio Carmeli and Teiko Heinosaari and Alessandro Toigo},
title = {Quantum random access codes and incompatibility of measurements},
journal = {Europhysics Letters}
}

@article{Bischof2019_PRL_Coherence,
  title = {Resource Theory of Coherence Based on Positive-Operator-Valued Measures},
  author = {Bischof, Felix and Kampermann, Hermann and Bru\ss{}, Dagmar},
  journal = {Phys. Rev. Lett.},
  volume = {123},
  year = {2019},
  publisher = {American Physical Society},
  doi = {10.1103/PhysRevLett.123.110402},
  url = {https://link.aps.org/doi/10.1103/PhysRevLett.123.110402}
}

@article{Bischof2021PhysRevA,
  title = {Quantifying coherence with respect to general quantum measurements},
  author = {Bischof, Felix and Kampermann, Hermann and Bru\ss{}, Dagmar},
  journal = {Phys. Rev. A},
  volume = {103},
  year = {2021},
  publisher = {American Physical Society},
  doi = {10.1103/PhysRevA.103.032429},
  url = {https://link.aps.org/doi/10.1103/PhysRevA.103.032429}
}

@article{Spekkens2009,
  title = {Preparation Contextuality Powers Parity-Oblivious Multiplexing},
  author = {Spekkens, Robert W. and Buzacott, D. H. and Keehn, A. J. and Toner, Ben and Pryde, G. J.},
  journal = {Phys. Rev. Lett.},
  volume = {102},
  issue = {1},
  pages = {010401},
  numpages = {4},
  year = {2009},
  month = {Jan},
  publisher = {American Physical Society},
  doi = {10.1103/PhysRevLett.102.010401},
  url = {https://link.aps.org/doi/10.1103/PhysRevLett.102.010401}
}

@article{Pan2019,
author={Pan, A. K.},
title={Revealing universal quantum contextuality through communication games},
journal={Scientific Reports},
year={2019},
month={Nov},
day={26},
volume={9},
number={1},
pages={17631},
issn={2045-2322},
doi={10.1038/s41598-019-53701-5},
url={https://doi.org/10.1038/s41598-019-53701-5}
}

@article{Patra2024,
  doi = {10.22331/q-2024-04-09-1315},
  url = {https://doi.org/10.22331/q-2024-04-09-1315},
  title = {Classical analogue of quantum superdense coding and communication advantage of a single quantum system},
  author = {Patra, Ram Krishna and Naik, Sahil Gopalkrishna and Lobo, Edwin Peter and Sen, Samrat and Guha, Tamal and Bhattacharya, Some Sankar and Alimuddin, Mir and Banik, Manik},
  journal = {{Quantum}},
  issn = {2521-327X},
  publisher = {{Verein zur F{\"{o}}rderung des Open Access Publizierens in den Quantenwissenschaften}},
  volume = {8},
  pages = {1315},
  month = apr,
  year = {2024}
}

@article{saha2023pra,
  title = {Measurement incompatibility and quantum advantage in communication},
  author = {Saha, Debashis and Das, Debarshi and Das, Arun Kumar and Bhattacharya, Bihalan and Majumdar, A. S.},
  journal = {Phys. Rev. A},
  volume = {107},
  issue = {6},
  pages = {062210},
  numpages = {14},
  year = {2023},
  month = {Jun},
  publisher = {American Physical Society},
  doi = {10.1103/PhysRevA.107.062210},
  url = {https://link.aps.org/doi/10.1103/PhysRevA.107.062210}
}

@article{plenioprl19,
  title = {Quantifying Operations with an Application to Coherence},
  author = {Theurer, Thomas and Egloff, Dario and Zhang, Lijian and Plenio, Martin B.},
  journal = {Phys. Rev. Lett.},
  volume = {122},
  issue = {19},
  pages = {190405},
  numpages = {6},
  year = {2019},
  month = {May},
  publisher = {American Physical Society},
  doi = {10.1103/PhysRevLett.122.190405},
  url = {https://link.aps.org/doi/10.1103/PhysRevLett.122.190405}
}

@article{ghorai2018,
  title = {Optimal quantum preparation contextuality in an $n$-bit parity-oblivious multiplexing task},
  author = {Ghorai, Shouvik and Pan, A. K.},
  journal = {Phys. Rev. A},
  volume = {98},
  issue = {3},
  pages = {032110},
  numpages = {8},
  year = {2018},
  month = {Sep},
  publisher = {American Physical Society},
  doi = {10.1103/PhysRevA.98.032110},
  url = {https://link.aps.org/doi/10.1103/PhysRevA.98.032110}
}

@article{chen_banik2024,
  title = {Quantum Advantage: A Single Qubit's Experimental Edge in Classical Data Storage},
  author = {Ding, Chen and Lobo, Edwin Peter and Alimuddin, Mir and Xu, Xiao-Yue and Zhang, Shuo and Banik, Manik and Bao, Wan-Su and Huang, He-Liang},
  journal = {Phys. Rev. Lett.},
  volume = {133},
  issue = {20},
  pages = {200201},
  numpages = {7},
  year = {2024},
  month = {Nov},
  publisher = {American Physical Society},
  doi = {10.1103/PhysRevLett.133.200201},
  url = {https://link.aps.org/doi/10.1103/PhysRevLett.133.200201}
}

@article{abhyoudai23,
  title = {Robust certification of unsharp instruments through sequential quantum advantages in a prepare-measure communication game},
  author = {S. S., Abhyoudai and Mukherjee, Sumit and Pan, A. K.},
  journal = {Phys. Rev. A},
  volume = {107},
  issue = {1},
  pages = {012411},
  numpages = {18},
  year = {2023},
  month = {Jan},
  publisher = {American Physical Society},
  doi = {10.1103/PhysRevA.107.012411},
  url = {https://link.aps.org/doi/10.1103/PhysRevA.107.012411}
}

@misc{aberg2006quantifyingsuperposition,
      title={Quantifying Superposition}, 
      author={Johan Aberg},
      year={2006},
      eprint={quant-ph/0612146},
      archivePrefix={arXiv},
      primaryClass={quant-ph},
      url={https://arxiv.org/abs/quant-ph/0612146}, 
}

@article{Frenkel2022entanglement,
  doi = {10.22331/q-2022-03-01-662},
  url = {https://doi.org/10.22331/q-2022-03-01-662},
  title = {On entanglement assistance to a noiseless classical channel},
  author = {Frenkel, P{\'{e}}ter E. and Weiner, Mih{\'{a}}ly},
  journal = {{Quantum}},
  issn = {2521-327X},
  publisher = {{Verein zur F{\"{o}}rderung des Open Access Publizierens in den Quantenwissenschaften}},
  volume = {6},
  pages = {662},
  month = mar,
  year = {2022}
}

@ARTICLE{Shannon1948,
  author={Shannon, C. E.},
  journal={The Bell System Technical Journal}, 
  title={A mathematical theory of communication}, 
  year={1948},
  volume={27},
  number={3},
  pages={379-423},
  doi={10.1002/j.1538-7305.1948.tb01338.x}}

@ARTICLE{Shannon1949,
  author={Shannon, C.E.},
  journal={Proceedings of the IRE}, 
  title={Communication in the Presence of Noise}, 
  year={1949},
  volume={37},
  number={1},
  pages={10-21},
  doi={10.1109/JRPROC.1949.232969}}

@article{Hartley1928Transmission,
  author  = {Hartley, R. V. L.},
  title   = {Transmission of Information},
  journal = {Bell System Technical Journal},
  year    = {1928},
  volume  = {7},
  number  = {3},
  pages   = {535--563},
  doi     = {10.1002/j.1538-7305.1928.tb01236.x},
  url     = {https://doi.org/10.1002/j.1538-7305.1928.tb01236.x},
  issn    = {0005-8580},
}

@book{Wilde2017,
  author    = {Wilde, Mark M.},
  title     = {Quantum Information Theory},
  edition   = {2nd},
  year      = {2017},
  publisher = {Cambridge University Press},
  address   = {Cambridge},
  doi       = {10.1017/9781316809976},
  isbn      = {9781107176164}
}

@book{NielsenChuang2010,
  author    = {Nielsen, Michael A. and Chuang, Isaac L.},
  title     = {Quantum Computation and Quantum Information: 10th Anniversary Edition},
  publisher = {Cambridge University Press},
  year      = {2010},
  address   = {Cambridge},
  isbn      = {9781107002173},
  doi       = {10.1017/CBO9780511976667}
}

@book{busch1996quantum,
  title={The Quantum Theory of Measurement},
  author={Busch, Paul and Lahti, Pekka and Mittelstaedt, Peter},
  year={1996},
  publisher={Springer Berlin Heidelberg},
  edition={2nd},
  volume={2},
  isbn={978-3-540-37205-9},
  doi={https://doi.org/10.1007/978-3-540-37205-9}
}

@article{Sarkar2024networkquantum,
  doi = {10.22331/q-2024-07-19-1419},
  url = {https://doi.org/10.22331/q-2024-07-19-1419},
  title = {Network quantum steering enables randomness certification without seed randomness},
  author = {Sarkar, Shubhayan},
  journal = {{Quantum}},
  issn = {2521-327X},
  publisher = {{Verein zur F{\"{o}}rderung des Open Access Publizierens in den Quantenwissenschaften}},
  volume = {8},
  pages = {1419},
  month = jul,
  year = {2024}
}

@article{RochiCarceller_2026,
doi = {10.1088/2058-9565/ae4d80},
url = {https://doi.org/10.1088/2058-9565/ae4d80},
year = {2026},
month = {mar},
publisher = {IOP Publishing},
volume = {11},
number = {2},
pages = {025020},
author = {Roch i Carceller, Carles and Tavakoli, Armin},
title = {The role of entanglement in energy-restricted communication and randomness generation},
journal = {Quantum Science and Technology},
}

@article{Bohr_2022,
  title = {Quantum vs Noncontextual Semi-Device-Independent Randomness Certification},
  author = {Roch i Carceller, Carles and Flatt, Kieran and Lee, Hanwool and Bae, Joonwoo and Brask, Jonatan Bohr},
  journal = {Phys. Rev. Lett.},
  volume = {129},
  issue = {5},
  pages = {050501},
  numpages = {6},
  year = {2022},
  month = {Jul},
  publisher = {American Physical Society},
  doi = {10.1103/PhysRevLett.129.050501},
  url = {https://link.aps.org/doi/10.1103/PhysRevLett.129.050501}
}

@article{SSarkarr2025APLQ,
    author = {Sarkar, Shubhayan},
    title = {Shared randomness allows violation of macroscopic realism using a single measurement},
    journal = {APL Quantum},
    volume = {2},
    number = {2},
    pages = {026128},
    year = {2025},
    month = {05},
    issn = {2835-0103},
    doi = {10.1063/5.0252123},
    url = {https://doi.org/10.1063/5.0252123}
}

@article{SSarkar2025PRA_Qresources,
  title = {Detecting quantum resources in a semi-device-independent framework},
  author = {Sarkar, Shubhayan and Datta, Chandan},
  journal = {Phys. Rev. A},
  volume = {111},
  issue = {4},
  pages = {L040402},
  numpages = {6},
  year = {2025},
  month = {Apr},
  publisher = {American Physical Society},
  doi = {10.1103/PhysRevA.111.L040402},
  url = {https://link.aps.org/doi/10.1103/PhysRevA.111.L040402}
}
\end{document}